\documentclass[a4paper,fleqn,usenatbib]{mnras}

\usepackage{newtxtext,newtxmath}

\usepackage[T1]{fontenc}
\usepackage{ae,aecompl}

\usepackage{graphicx}	
\usepackage{amsmath}	
\usepackage{amssymb}	
\usepackage{natbib}
\usepackage{indentfirst}
\usepackage{adjustbox}
\usepackage{supertabular,booktabs}
\usepackage{rotating}
\title[Interstellar Scintillation of an Extreme Scintillator: PKS\,B1144$-$379]{Interstellar Scintillation of an Extreme Scintillator: PKS\,B1144$-$379}

\author[N. M. M. Said et al.]{
N. M. M. Said,$^{1,2}$\thanks{E-mail: noor.mdsaid@utas.edu.au}
S. P. Ellingsen,$^{1}$
H. E. Bignall, $^{2}$
S. Shabala,$^{1}$
J. N. McCallum,$^{1}$ \newauthor \ 
C. Reynolds,$^{2}$
\\
$^{1}$School of Natural Sciences, University of Tasmania, Private Bag 37, Hobart, TAS 7001, Australia\\
$^{2}$CSIRO Astronomy and Space Science, P.O. Box 1130, Bentley, WA 6102, Australia\\
}

\date{Accepted XXX. Received YYY; in original form ZZZ}

\pubyear{2020}

\begin{document}
\label{firstpage}
\pagerange{\pageref{firstpage}--\pageref{lastpage}}
\maketitle

\begin{abstract}
The University of Tasmania Ceduna radio telescope has been used to investigate rapid variability in the radio flux density of the BL Lac object PKS\,B1144$-$379 at 6.7 GHz. High-cadence monitoring of this extreme scintillator was carried out over a period of approximately nine years, between 2003 and 2011. We have used structure functions created from the intensity time series to determine the characteristic timescale of the variability. The characteristic timescale is consistently observed to increase during certain periods of each year, demonstrating the annual cycle expected for scintillation through an interstellar scattering screen. The best-fitting annual cycle model for each year suggests that the scintillation pattern has an anisotropic structure and that the upper limit of its scattering screen is at a distance of $\sim$0.84 kpc. Higher anisotropy in some of the annual cycle fits suggests that changes in the intrinsic source structure might be influencing the variability timescale. We found a prominent annual cycle is only present in the variability timescale for certain years where other evidence suggests that the core is compact. From our measurements we calculated that the core angular size varied between 5.65--15.90 $\mu$as (0.05--0.13 pc). The core component was found to be at its most compact during two flares in the total flux density, which were observed in 2005 and 2008. We conclude that the long-term variability in the radio flux density of PKS\,B1144$-$379 is due to intrinsic changes in the source and that these affect our ability to measure an annual cycle in its variability time scale.
\end{abstract}

\begin{keywords}
galaxies: ISM, BL Lacertae objects: PKS\,B1144$-$379
\end{keywords}



\section{Introduction}

Active galactic nuclei (AGN) exhibit variability in their emission at frequencies across the entire electromagnetic spectrum. The measured flux density of these sources is known to vary on a range of timescales from a few days up to several months. The rapid (timescales of less than a day), apparent variability of these compact extragalactic radio sources at radio wavelengths is known as intraday variability (IDV). IDV of AGN at centimetre wavelengths was discovered in the 1980s \citep{Wagner-Witzel-1995,Heeschen-et-al-1987}.  In general, there are two possible interpretations of this phenomenon; it can either be due to an intrinsic cause where the AGN itself is changing or evolving, or have an extrinsic origin which involves a propagation effect along the line of sight, known as interstellar scintillation (ISS). An intrinsic cause of radio IDV is problematic as it requires the size of the emitting region to be very small \citep{Qian-et-al-1991}, brightness temperatures that significantly violate the inverse Compton limit \citep[$\sim 10^{12}$ K;][]{Kellermann-Pauliny-1969}, and unrealistically high Doppler beaming factors \citep{Kedziora-Chudzer-et-al-1997}.  ISS is an interference phenomenon seen in radio waves, caused by an inhomogeneous ionized medium between the source and observer \citep[e.g.][]{Rickett1990,Narayan1992,Jauncey-et-al-2016}. \par

There are two compelling pieces of evidence that it is ISS which causes radio IDV; annual modulation in the timescale of variability over the course of the year and a time delay in the variability pattern observed at two widely spaced radio telescopes. Annual modulation in characteristic timescale of variability arises due to the Earth's orbital motion changing the relative velocity between the observer and the scattering medium. Such annual modulations have been detected toward a number of sources, for example, J1819+3845 \citep{Dennett-Bruyn-2001, Dennett-Bruyn-2003}, QSO B0917+624 \citep{Jauncey-Macquart-2001,Rickett-et-al-2001,Fuhrmann-et-al-2002}, PKS\,B1257$-$326 \citep{Bignall-et-al-2003}, PKS\,B1519$-$273 \citep{Jauncey-et-al-2003,Carter-at-al-2009}, PKS\,B1622$-$253 \citep{Carter-at-al-2009}, S5\,0716$+$714 \citep{Liu-et-al-2012}, 0925$+$504 (\citep{Liu-Liu-2015,Liu-et-al-2017}), S4\,0954$+$65 \citep{Marchili-et-al-2012}, 1156$+$295 \citep{Liu-et-al-2013}, J1128$+$5925 \citep{Gabanyi-et-al-2007a,Gabanyi-et-al-2007b} and PKS\,1322$-$110 \citep{Bignall-et-al-2019}. Measurement of a time delay in the variability pattern between two widely spaced radio telescopes is feasible if the variability timescale is significantly shorter than the duration of the observing interval. \citet{Bignall-et-al-2006} reported time delays of up to 8 minutes in the centimetre wavelength variability pattern of the intrahour variable (IHV) scintillating quasar (type I radio IDV) PKS\,B1257$-$326 observed between the Very Large Array (VLA) and the Australia Telescope Compact Array (ATCA) on 3 separate epochs. \citet{Jauncey-et-al-2000} measured a time delay of around 2 mins between the ATCA and VLA for PKS\,B0405$-$385. Similarly, for J1819$+$3845, \citet{Dennett-Thorpe-de-Bruyn-2000,Dennett-Bruyn-2003} observed using the Westerbork Synthesis Radio Telescope (WSRT) and the VLA, finding a clear time delay of 2 mins between the pattern arrival times at each telescope which changed over the course of the observations. No clear annual cycle has been detected for the source PKS\,B0405$-$385 due to its episodic IHV behaviour \citep{Kedziora-Chudzer-2006}, and \citet{Fuhrmann-et-al-2002} reported the source B0917+624 had ceased its variability. The episodic behaviour of scintillation could be due to changes either in source structure or in the intervening screen properties. Observations of ISS can be used as a probe to study microarcsecond-scale source structure, as well as the properties of turbulence in the local Galactic ISM through the technique of ``Earth Orbital Synthesis'' \citep{Macquart-Jauncey-2002}. The high resolution capability of ISS exceeds that of Very Long Baseline Interferometry (VLBI) and possibly even that of space-based VLBI.\par

PKS\,B1144$-$379 ($11^{h}47^{m}01.4^{s}$, $-38^{\circ} 12^{\arcmin} 11^{\arcsec}$ J2000) is classified as a BL-Lac object due to its variability in optical, infrared and radio wavelengths \citep{Nicolson-et-al-1979,Stickel-et-al-1989}. This source is also listed as a quasar by \citet{Veron-Cetty-2006}. It is at a redshift $z$ =1.048 \citep{Stickel-et-al-1989} and was classified as a radio variable in the Parkes 2700 MHz survey \citep{Bolton-Shimmins-1973}. It is known to exhibit both long-term and short-term flux density variations (timescale of a few days; type II radio IDV) at centimetre wavelengths. Large amplitude radio IDV in both total and polarized flux density was first identified in this source by \cite{Kedziora-Chudzer-2001a,Kedziora-Chudzer-2001b}, and this was the reason for this object being included in the long-term 6.7-GHz monitoring project COSMIC (Continuous Single Dish Monitoring of IDV at Ceduna) using the University of Tasmania's 30-m Ceduna radio telescope \citep{McCulloch-et-al-2005,Carter-at-al-2009}. The long-term variability (months to years) of PKS\,B1144$-$379 suggests that this source also exhibits substantial intrinsic evolution. The VLBI image of this source from the TANAMI project carried out by \citet{Ojha-et-al-2010} shows a strong compact core and clear jet oriented south-east, with significant emission at about 30 mas from the core. This bright, rapidly variable radio source was not detected by the Energetic Gamma Ray Experiment Telescope (EGRET) but has been detected by the Large Area Telescope (LAT) \citep{Abdo-et-al-2009}. An investigation of the radio emission in this source was conducted by \citet{Turner-et-al-2012}. They observed PKS\,B1144$-$379 at 4 frequencies ranging from 1.5 to 15 GHz with the VLA and Ceduna 30-m telescopes. From their observations, they estimated the source angular size to be 20--40 $\mu$as (or 0.15--0.3 pc) and the inferred brightness temperature, assuming that the observed variations are due to scintillation, is 6.2x$10^{12}$ K at 4.9 GHz with approximately 10$\%$ of the total flux density in the scintillating component. PKS\,B1144$-$379 can be classified as an extreme scintillator based on its unusually large amplitude intraday variability, with modulation index in the range 5--18$\%$. Only 3.6$\%$ of sources in the 5~GHz MASIV VLA Survey of 475 compact radio sources had modulation index > 5$\%$, and 0.4$\%$ showed modulation index > 15$\%$ in observations over 3--4 days \citep{Lovell-et-al-2008}. Therefore, analysis of long-term monitoring of this source can improve our understanding of the origin of turbulent structures responsible for rare, extreme ISS \citep[e.g.][]{Walker-et-al-2017}. \par

In this paper, we use COSMIC data taken between 2003 and 2011 to search for evidence of annual modulation in the timescale of the centimetre wavelength radio emission from PKS\,B1144$-$379 and study the physical properties of the source and the scattering screen. In order to determine the characteristic timescale of the variability, which shows quasi-periodicity of around 3--4 days, observations with a duration greater than a week are needed. Therefore, a long-term, high-cadence monitoring programme, such as COSMIC, is required.  In \S 2 we discuss the observations, calibration and methods used to estimate the characteristic timescale of intensity changes from the Structure Function (SF). In \S 3 we investigate the evidence for an annual cycle in the variability timescale and compare it to the expectations of both isotropic and anisotropic scattering screen models. We investigate the reliability of the SF method through Monte Carlo simulations in \S 4 and the results and discussions of annual cycle fitting for the years 2003 to 2011 and the fitted parameters are presented in \S 5 and 6. We summarise our main conclusions in \S 7. \par

\section{Observations and data analysis}

In this section, we give details of the COSMIC observing project and the associated data calibration process. We also demonstrate the estimation of characteristic timescales using the structure function (SF) method.

\subsection{Observations and Calibration}

In this paper, we use data from the COSMIC (Continuous Single-dish Monitoring of Intraday Variability at Ceduna) observing project to investigate the variable radio source PKS\,B1144$-$379. The COSMIC project was designed to provide high-cadence monitoring of the flux density variations of several strong sources of radio IDV using the 30-m Ceduna radio telescope. The observations started in early 2003 and continued until 2013. The Ceduna 30-m telescope was equipped with an uncooled receiver with two orthogonal circular polarizations operating over the frequency range of 6.4--6.9 GHz (centred at 6.7 GHz, covering a bandwidth of approximately 500 MHz). This frequency was selected as it corresponds to the approximate transition frequency between weak and strong scattering for extragalactic sources at high Galactic latitude and the transition frequency is where the variability amplitude is largest \citep{Walker1998,Walker2001}. More details about the COSMIC project, including the procedures used for the observations and an outline of the data reduction, can be found in \citet{McCulloch-et-al-2005}. The methods for reduction of COSMIC data were refined by \citet{Blanchard2013} and we have utilised those methods here.  \par

Here we give a brief summary of the observational method.  A single measurement of the flux density of a source is obtained from two pairs of orthogonal scans. The first pair of scans has one taken with increasing Right Ascension with constant Declination, followed by another taken with decreasing Right Ascension.  The second pair holds the Right Ascension constant and scans first in increasing, then decreasing Declination. All the scans are initially centred on the nominal source position, but the online observing system estimates pointing offsets from the orthogonal scans and updates the position for future observations.  Updated pointing offsets are only utilised if less than four hours old, otherwise, the nominal pointing position is used.  Scans are saved in FITS format files, along with appropriate metadata. Obtaining a flux density measurement from the two pairs of scans is a multi-step process, with each group of four scans toward a single source considered separately. The amplitude data is first normalised using the receiver noise diode and the amplitude as a function of position on the sky is then fitted with a Gaussian profile and a polynomial baseline. The next step is to average together the amplitude data for the pair of Right Ascension (constant Declination) scans and similarly for the pair of Declination scans.  The two averaged scans are then compared for consistency and a group of four scans is only accepted for further processing if the amplitude for the two average scans is the same to within 10 $\%$. The next step is to estimate the pointing offset between the observed and actual source position and correct the measured amplitude for that offset.  The observed amplitude for the Right Ascension average scan will be reduced by a Declination offset between the observed and measured source position.   We measure this offset from the position of the peak in the Declination average scan, calculate the appropriate correction factor using the known shape of the beam profile and apply it to the Right Ascension scan. The same process is then repeated for the Declination average scan using the offset measured in the Right Ascension average scan. The two pointing-corrected averaged scans are then themselves averaged. Finally, a gain-elevation correction is made based on a gain curve calculated from one year of observations of PKS\,B1934$-$638, a southern calibrator known to be flux density stable, with an assumed flux density of 3.9383 Jy at 6.7 GHz \citep[with an uncertainty of 2$\%$;][]{Reynolds1994}. The gain-elevation curve is applied to correct for small changes in the gain of the antenna with elevation. \par

To reduce the amount of observing time lost to antenna slewing the COSMIC project split its target sources into two groups, those that passed north and south of the zenith. Each group of sources was typically observed for periods of 10--14 days at a time, alternating between the two groups. PKS\,B1144$-$379 was an exception and from early 2004 was included in both groups (meaning near-continuous monitoring). Most sources (including calibrators) were only observed in one group and for northern sources, 3C 227 (assumed flux density of 1.9885 Jy) was the flux density calibrator \citep[with an uncertainty of 2$\%$;][]{Baars-et-al-1977}.  The amplitude scale for the data was converted into flux density in Jansky (Jy) by calculating the noise diode amplitude through comparison with either PKS\,B1934$-$638 (southern calibrator) or 3C 227 (northern calibrator). The COSMIC project made observations of the target and calibrator sources only when they were at elevations greater than 10 degrees and the measured gain variations of the antenna are less than 5$\%$ \citep{McCulloch-et-al-2005}. It was found that after scaling using the gain-elevation correction, the measured root mean square (RMS) of the flux density for three calibrator sources (PKS\,B1934$-$638, PKS\,B0945$+$076 (3C227) and PKS\,B1921$-$293) has a term that is some fraction of the source flux density. Fitting the observed RMS of these calibrator sources with the quadrature function gives a constant error term of 29 mJy and a fractional term of 0.84$\%$ of the source flux density (in Jy). All the scintillating sources observed in the COSMIC programme have  mean flux densities at 6.7 GHz that are typically in the range 1--4 Jy. \par

\begin{figure}
	\includegraphics[width=\columnwidth]{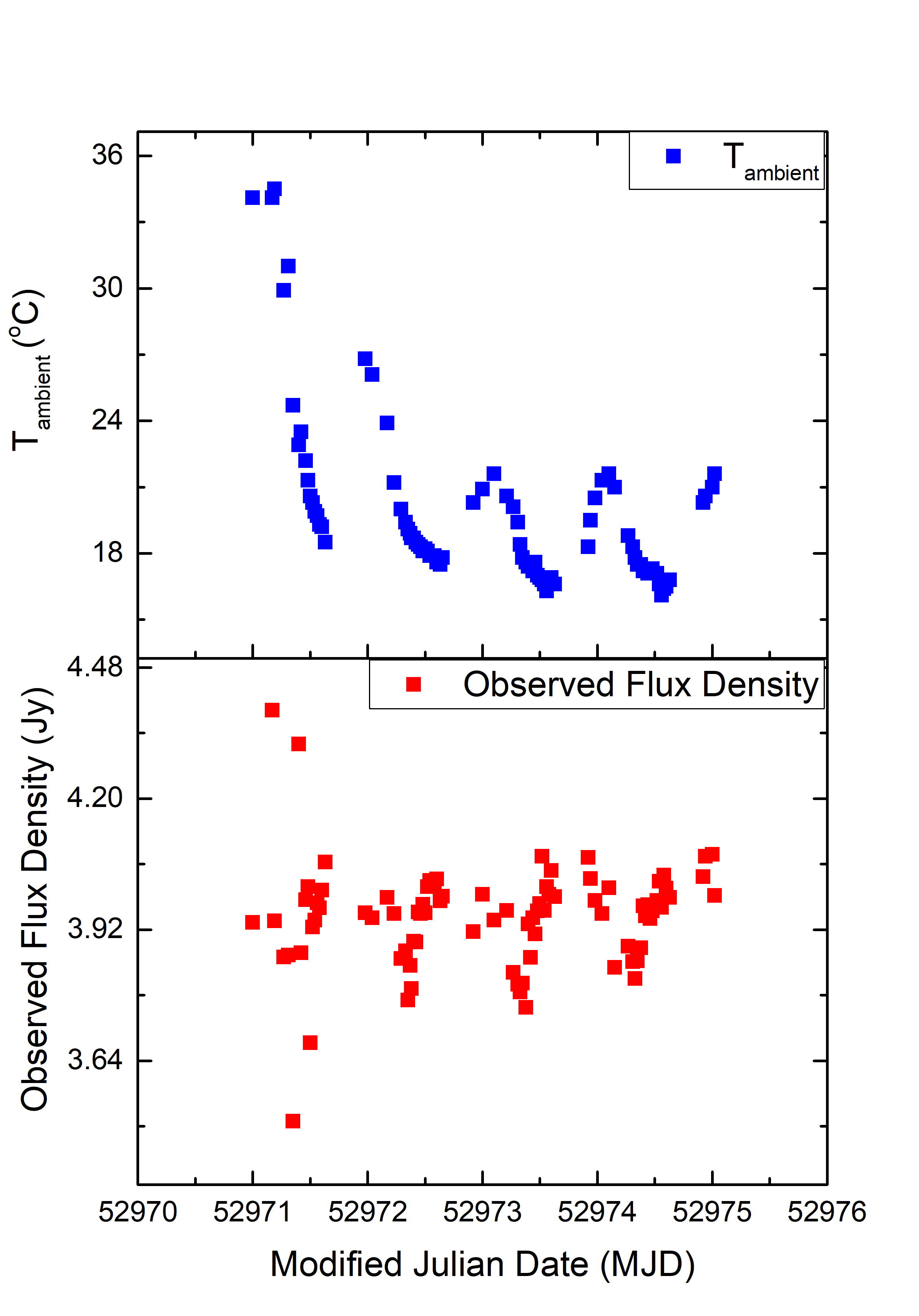}
	\caption{Top: External temperature at the Ceduna observatory over a 6 day period in November 2003. Bottom: Apparent diurnal variations in the amplitude of the flux density calibrator PKS\,B1934$-$638 over the same period.}
	\label{fig:temperature}
\end{figure}

After applying the calibration procedures developed by \citet{Blanchard2013}, the examination of the flux density calibrators shows RMS residual variations of approximately 3$\%$. These variations have been identified as systematic errors due to a small temperature dependence of the noise diode amplitude (which is not in a temperature-controlled environment), and this, in turn, creates systematic errors in the gain-elevation curve. In the previous analysis using COSMIC data, the amplitude of the noise diode was assumed to be constant and was determined from the assumed flux density of the calibrator source (a constant value which does not vary with time). However, the temperature dependence of the noise diode amplitude means that daily temperature variations result in apparent diurnal variations in the flux density of the calibrator. Figure~\ref{fig:temperature} shows the variations in the external temperature and the flux density of PKS\,B1934$-$638 measured assuming a constant amplitude for the noise diode, over a 6 day period. Diurnal variations in the apparent flux density of the calibrator are clearly visible in the lower panel of Figure~\ref{fig:temperature}. We investigated a number of different approaches to correct for the temperature dependence of the noise diode, but were unable to find any method which reliably reduced the systematic variations below 3$\%$.  While we have good data on the external air temperature during the observations, the noise diode is located in the receiver room, where the ambient temperature differs from the external temperature \citep{Senkbeil-et-al-2008} and changes in the ambient temperature of the receiver room lag external temperature variations, but not by a constant nor predictable amount. The temperature dependence of the noise diode affects the flux density measurements of both the calibrator and the target sources. Fortunately, the amplitude of the typical ISS-induced variations in the target sources is much larger than 3$\%$, however, it is important during the data analysis that any methods applied to the target source must be applied to the calibrator as well. In this way, the effect of any systematic variations in the flux density of the target source can be estimated. It is vital to properly characterize the systematic variations in the flux density of the calibrator and target source, as it may impact the measurement of the characteristic timescale of variations, and through that subsequent time series analysis of the target source. \par

\begin{figure}
	\includegraphics[width=\columnwidth]{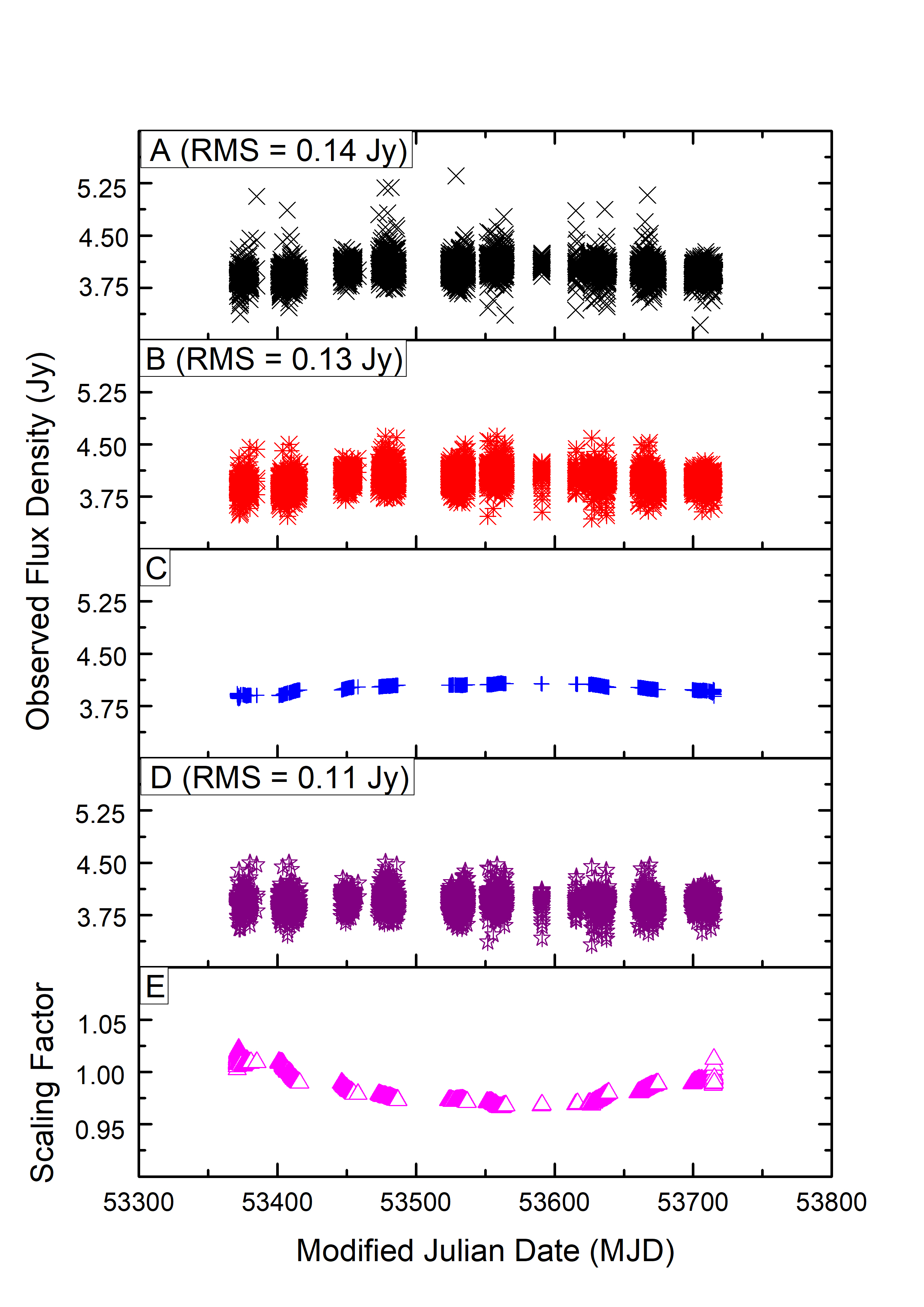}
	\caption{Light curves of the southern calibrator PKS\,B1934$-$638 during year 2005. Panels show all the procedures applied to the flux density of the calibrator to reduce the systematic errors. Panel A: raw observed flux density; B: observed flux density after eliminating 5$\sigma$ outliers; C: smoothed observed flux density; D: final estimate of the observed flux density; and E: measured scaling factor.}
	\label{fig:scale_factor}
\end{figure}

\begin{figure}
	\includegraphics[width=\columnwidth]{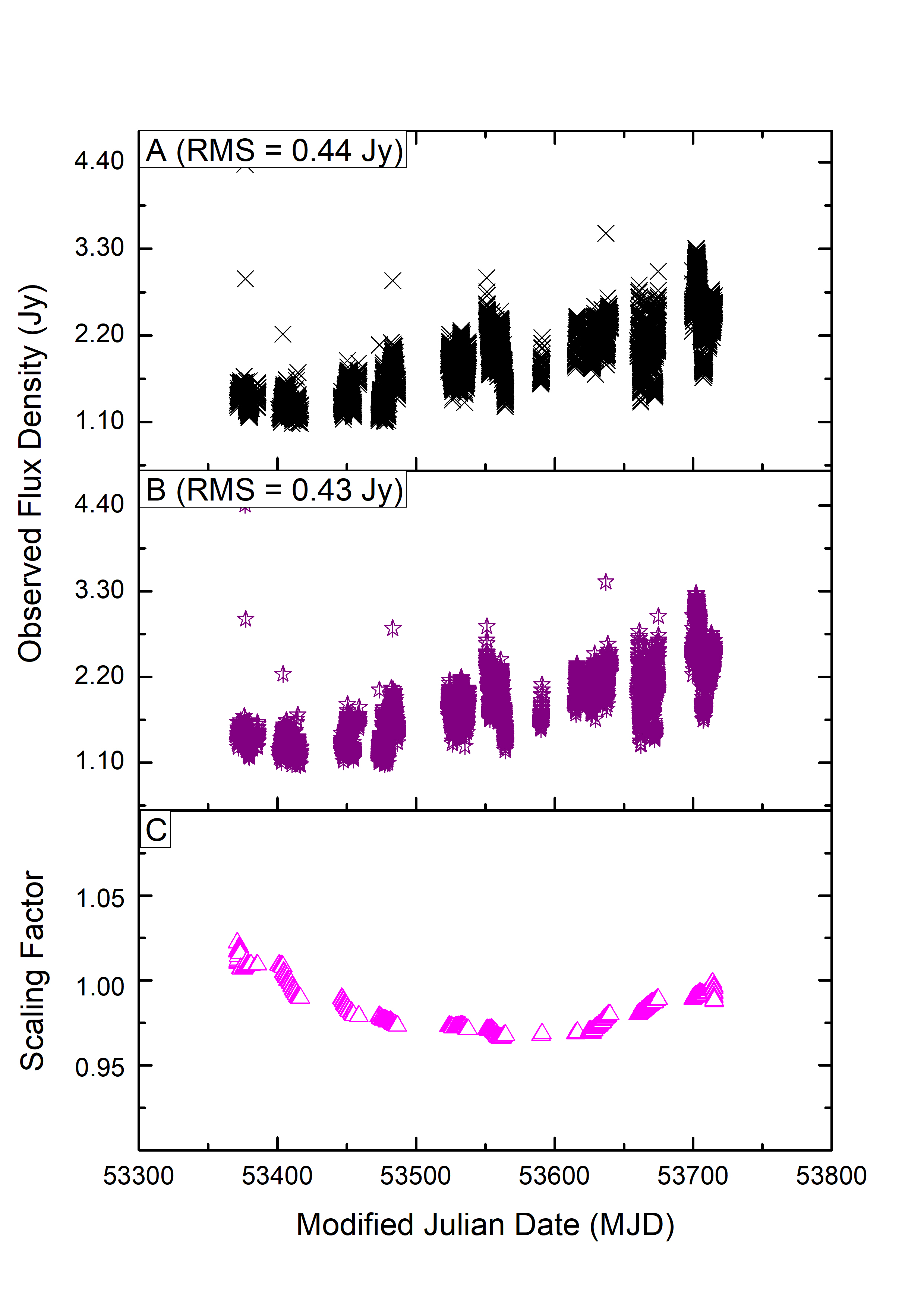}
	\caption{Light curves of the target source PKS\,B1144$-$379 during year 2005. Panel A: raw observed flux density; B: final estimate of the observed flux density of the target source and C: measured scaling factor of the target source.}
	\label{fig:scale_factor_target}
\end{figure}

We filtered the flux density data for the calibrators PKS\,B1934$-$638 (southern calibrator) and 3C227 (northern calibrator) in order to remove extreme outliers. The mean and standard deviation ($\sigma$) of the flux density for one year of observations were calculated and any data that differed from the mean by more than 5$\sigma$ were eliminated. Removing the extreme outliers reduced the standard deviation in the flux density of the calibrators by $\sim$7$\%$. We then applied a smoothing procedure using a Simple Moving Average (SMA) technique to the flux density of both calibrators. This technique is a type of Finite Impulse Response (FIR) filter which is applied to a set of data points by creating an average of different subsets of the full dataset \citep{Azami-et-al-2012}. The observations of the flux density calibrators for each year were smoothed with 20$\%$ of the sample size of the calibrator data as the weighted average. The next step is to measure the scale factor for the amplitude of the noise diode. The scale factor was determined from the ratio of the smoothed flux densities and the assumed flux density of the calibrators (3.9383 and 1.9885 Jy for the southern and northern calibrators respectively at a frequency of 6.7 GHz). Figure~\ref{fig:scale_factor} shows the flux density data for the calibrator PKS\,B1934$-$638 (southern calibrator) after it has undergone all the procedures to reduce the systematic errors. After applying all the procedures, the standard deviation of flux density of the calibrator was reduced by $\sim$22$\%$. As mentioned before, it is important during the data analysis that any methods applied to the target source must be applied to the calibrator as well. We, therefore, measured the noise-diode scaling factor for the target source from the scale factor of the calibrator. This procedure was done by linearly interpolating scale factor of the target source from the scale factor of the calibrator for the particular period of the observing epoch (MJD). Figure~\ref{fig:scale_factor_target} illustrates the flux density of the target source PKS\,B1144$-$379 prior and after we applied the scale factor. The scale factors for both calibrators and target sources are less than 5$\%$. The final estimate of the flux density for the calibrators and target source was obtained by multiplying the data with the scale factors. Prior to undertaking time-series analysis, the data for each source were split into ``blocks'', which typically have a duration of $\sim$14 days. The blocks generally correspond to the periods of time for which we observed either the southern or northern source groups. For most blocks, we have flux density data for the calibrator and target sources at least once per hour for the times that they were above 10 degrees elevation for the Ceduna antenna and there are no breaks in the data for any block of more than 5 days. Any outliers in each block of data for target sources were removed before performing the time-series analysis. A previous analysis of COSMIC data used a different approach to removing the systemic variations that involved smoothing and resampling the data \citep[e.g.][]{Carter-at-al-2009}. \citet{Carter-at-al-2009} combined the use of discrete autocorrelation functions and the related power spectral density to estimate the characteristic timescale of the variations. This approach was effective, however, it does intrinsically low-pass filter the timescale of variations. It was therefore not possible to reliably determine if there are multiple variability timescales present in the data with the method of \citet{Carter-at-al-2009}. In the present paper, we analyse the data using a different method and without smoothing or resampling, as we have reduced the level of systematic variations (3$\%$) well below the level of radio IDV observed for PKS\,B1144$-$379 which has high modulation index of 5--18$\%$. Our method potentially allows the detection of multiple variability timescales present in the data, although for the present paper we define a single characteristic timescale.  \par

\subsection{Estimating the characteristic timescale using structure functions}

\begin{figure*}
	\centering
	\includegraphics[width=0.9\textwidth]{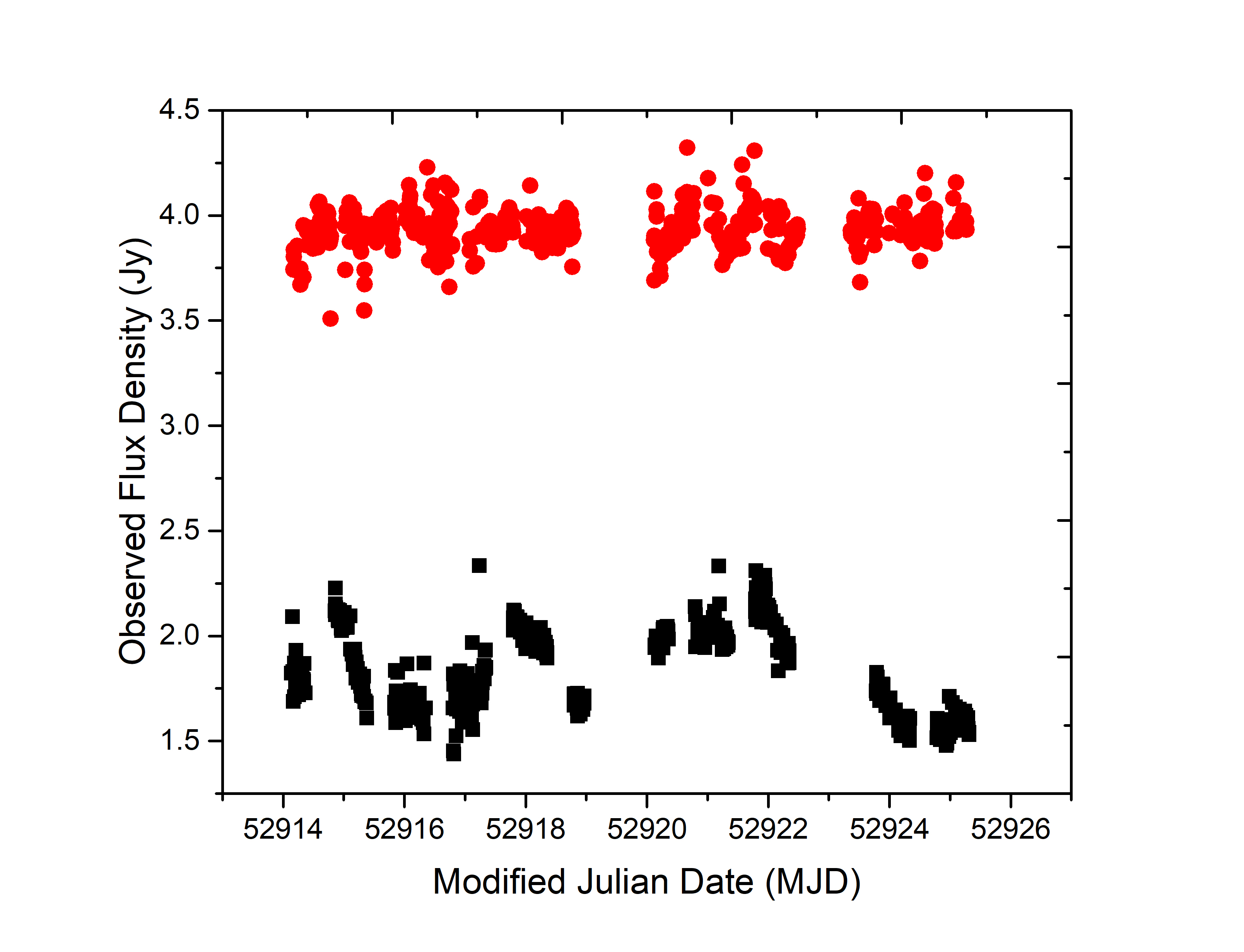}
	\caption{Light curves of the southern calibrator PKS\,B1934$-$638 (red) and target source PKS\,B1144$-$379 (black) for a 12-day observing block in October 2003.}
	\label{fig:lightcurve}
\end{figure*}

There are several methods that have been used to measure the variability timescale of radio IDV. For example \citet{Simonetti-et-al-1985} and \citet{Dennett-Bruyn-2003} use the structure function (SF) of the measured time series to define the characteristic timescale, while \citet{Rickett-et-al-1995,Rickett-et-al-2002} estimate the variability timescale from the half-width at half maximum of the autocorrelation function (ACF), \citet{Kedziora-Chudzer-et-al-1997} used the mean peak-to-peak time, \citet{Jauncey-Macquart-2001} adopted the peak-to-trough time and \citet{Gabanyi-et-al-2007b} estimated the timescale using 3 different methods; average peak-to-trough of the light curve, SF and the ACF. Due to the inherently stochastic nature of the flux density variations in AGN produced by ISS, it is difficult to define a characteristic timescale by simply examining the light curve (see for example Figure~\ref{fig:lightcurve}). For our analysis we chose to use the SF of the flux density variations (the light curve) to estimate the characteristic timescale. Prior to calculating the SF, the amplitude of the light curves of the calibrator and target source were normalised by dividing all the flux densities for each source light curve by the mean flux density of that source for the specific observing block. This ensures that the amplitudes of the SF are related to the fractional variation and can be directly compared for the calibrator and target sources. The SF method was first introduced by \citet{Simonetti-et-al-1985} and further described by \citet{Sprangler-et-al-1989}. The SF is defined to be the average squared difference between points of a data series, as a function of lag. The SF $D(\tau)$ is computed in a similar manner to the Discrete Correlation Function (DCF) by taking the difference between every point in the data series with every other point. The individual differences are then sorted according to the time between the differenced data points and binned accordingly  \par

\begin{equation}
D(\tau) = \frac{1}{N_{\tau}}\Sigma_{\textnormal{j,k}}(S_{\textnormal{j}} - S_{\textnormal{k}})^{2}
\label{eq:SF}
\end{equation}

\noindent
where $S_{\textnormal{j}}$ and $S_{\textnormal{k}}$ are flux density measurements and $N_{\tau}$ is the number of pairs of flux densities with a time lag of $\tau$. \par

\begin{figure*}
	\centering
	\includegraphics[width=\textwidth]{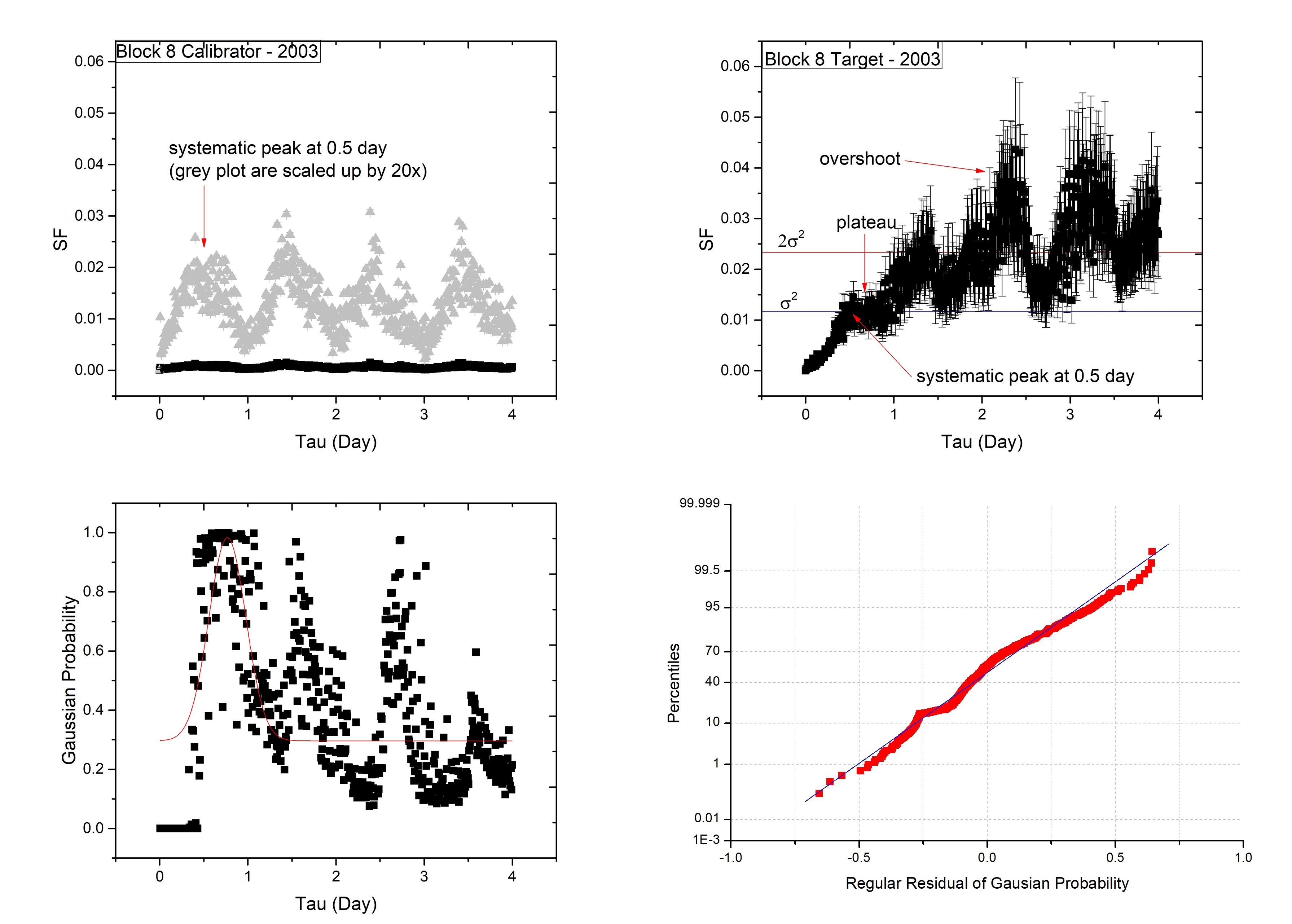}
	\caption{The normalised Structure Functions of the southern calibrator PKS\,1934$-$638 (upper-left) and target source PKS\,B1144$-$379 (upper-right) are shown on the same scale for a 12 day observing block in October 2003 (see Fig.~\ref{fig:lightcurve}).  In order to be able to clearly see the effect of the systematic variations in the calibrator (the grey points in the upper-left panel), we show the data scaled up by a factor of 20 times.  The bottom-left panel shows the Gaussian probability of SF target source with its peak and FWHM correspond to the estimated characteristic timescale ($\tau_{\textnormal{char}}$) and its uncertainty (red curve represents the Gaussian fitting), and the Quantile-Quantile plot of the Gaussian probability of SF target source is shown in the bottom-right panel. The horizontal red and blue lines in the SF of target source represent the saturated value ($2\sigma^{2}$) and variance level.}
	\label{fig:SF}
\end{figure*}

The SF of a source exhibiting variation due to ISS is usually described by a power law rising from a value of zero at zero lag to the peak plateau level, at a given lag. The amplitude of the SF is linked to the RMS difference at a given lag by twice the variance $(2\sigma^{2})$. In our case, the peak amplitude of the SF of all blocks exceeds the twice variance level (``overshoot'') and this can potentially lead to overestimating the variability timescale. The upper right-hand panel in Figure~\ref{fig:SF} shows the structure function calculated for the PKS\,B1144$-$379 light curve shown in Figure~\ref{fig:lightcurve}. The SF in the upper right-hand panel of Figure~\ref{fig:SF} shows a plateau commencing at around a lag of 0.5--1 day before an ``overshoot'' which peaks around 2.3 days.  The overshoot is linked to quasi-periodic behaviour in the variability light curve and is often observed for radio IDV sources \citep[e.g.][]{Rickett-et-al-2002}. To better estimate the characteristic timescale, instead of trying to fit to determine the amplitude of the SF plateau, we use the theoretical estimate of the saturation level (plateau) of twice the variance (red line in upper right-hand panel of Fig.~\ref{fig:SF}). We have used the time lag at half of the saturation value for our estimate of the characteristic timescale (blue line in upper right-hand panel of Fig.~\ref{fig:SF}). The estimated characteristic timescale corresponds to the length of the average peak-to-peak modulation divided by 4.  \par

As previously mentioned the Ceduna data suffers from $\sim3\%$ systematic variations which are predominantly diurnal and manifest as a peak at 0.5 day lag (see grey points in upper left-hand panel of Fig.~\ref{fig:SF}). Therefore, prior to estimating the characteristic timescale for target sources, we subtracted the SF of the calibrator (for which we assume the variations are residual systematic gain and diurnal variations) from the SF of target source. The resulting SF we assume to be a ``clean'' SF for the target source. We take the ``clean'' SF of the target source and performed a linear fit of the SF data with an amplitude less than twice the variance (saturation level).  The intercept between this straight line fit and the variance (half the saturation level) we use as our initial estimate of the characteristic timescale. However, because there is intrinsic scatter in the SF the ``final'' (best) estimate of the characteristic timescale and its uncertainty is obtained by calculating the Gaussian probability distribution given by equation~\ref{eq:gaussian probability}, which estimates the probability of a particular lag $\tau$ corresponding to the point at which the SF is equal to the light-curve variance.  The bottom-left panel of Figure~\ref{fig:SF} shows the Gaussian probability distribution for the SF in the upper right-hand panel.  We take the peak of the Gaussian probability distribution to be the characteristic timescale and its 1$\sigma$ uncertainty is computed from the Full Width Half Maximum (FWHM) of a Gaussian fitted to the Gaussian probability distribution. Quantile-Quantile (Q-Q) plots (lower right-hand panel) were constructed in order to quantify the reliability of Gaussian fitting using a Gaussian probability distribution (Fig.~\ref{fig:SF}). A Q-Q plot is another graphic method for testing whether a dataset follows a given distribution, in which our case is the normal probability distribution. It differs from the probability plot in that it shows observed and expected values instead of percentages on the X and Y axes. If all the scatter points are close to the reference line, we can say that the dataset follows the given distribution.    \par

\begin{equation}
\mbox{Probability}(\tau) = 
\exp \left \{\dfrac{-(SF(\tau)-(2\sigma^{2}/2))^{2}}{2\sigma_{\textnormal{SF}}^2(\tau)} \right \}
\label{eq:gaussian probability}
\end{equation}

\noindent
where SF$(\tau)$ is the initial characteristic timescale and the twice variance $(2\sigma^{2})$ is the saturation level of the SF. \par

We adopted the method of \citet{Rickett-et-al-2002} to quantify the uncertainty in the estimated SF for each lag $\tau$ using their formula:
  
\begin{equation}
 \sigma(\tau)_{\textnormal{ACF}}= \left \{1-\exp \left \{\dfrac{[-2(\tau/1.2\tau_{\textnormal{char}})^{2}]1.5\tau_{\textnormal{char}}}{T_{\textnormal{obs}}} \right \} \right \} ^{1/2}
 \label{eq:SF uncertainty}
\end{equation}
 
\noindent
where $\tau_{\textnormal{char}}$ is the estimated characteristic timescale in question and $T_{\textnormal{obs}}$ is the duration of the observation. This error function increases rapidly from zero at zero lag to a saturation at the lag where the SF reaches twice the variance. The uncertainty of the SF measured using equation~\ref{eq:SF uncertainty} assumes a Gaussian form for the Autocorrelation Function (ACF). The values of the measured uncertainty were then multiplied with the values of the measured SF in order to scale it appropriately for the SF method. The uncertainties computed using equation~\ref{eq:SF uncertainty} are not in the form of white noise, but are correlated across adjacent time lags. The magnitude of the uncertainty in the value of the SF for a particular lag $\tau$ depends on the estimated characteristic timescale, which is in turn derived from the noisy SF \citep{McCallum2009}. Limited sampling of a stochastic process is the dominant contribution to the uncertainty in the determination of the characteristic timescale. Generally speaking, the uncertainty decreases with the number of independent samples of the scintillation pattern, or ``scints'' observed \citep{Bignall-et-al-2003}. \par

\subsection{Measurement of variability amplitude}

Limited sampling is also the dominant source of uncertainty in determining the amplitude of variability. One measure of flux density variability in a light curve is the modulation index and this has been used by various authors to characterise AGN variability \citep[e.g.][]{Rickett-et-al-2001,Carter-at-al-2009,Marchili-et-al-2012}. The modulation index, $\textit{m}$ = $\sigma_{\textnormal{S}}$/$\bar{S}$, is defined as the RMS variation $\sigma_{\textnormal{S}}$ normalised by the average flux density of the source $\bar{S}$. Although the modulation index has the advantage of being non-negative and more robust against outliers, it still consists of a convolution of intrinsic source variation and observational uncertainties. The large modulation index could be an indication of either a strongly variable source or a faint source with high uncertainties in the measurement of individual flux density. Due to this effect, the accurate measurement of modulation index requires one to take into account the measurement errors and the uncertainty in the modulation index caused by the finite number of flux density measurements. To examine the accuracy of our measurement of modulation index, we decided to measure the intrinsic modulation index, $\bar{m}$ = $\sigma_{\textnormal{o}}$/$\bar{S}_{\textnormal{o}}$, which is the intrinsic RMS variation $\sigma_{\textnormal{o}}$ normalised by the intrinsic average flux density of the source $\bar{S}_{\textnormal{o}}$. This method was introduced by \citet{Richards-et-al-2011} where they used a likelihood analysis to obtain the intrinsic modulation index as well as its uncertainty. The term ``intrinsic'' is used to denote flux densities and variations as would be observed with a perfectly uniform sampling of adequate cadence and zero observational error. Thus, $\bar{m}$ is a measure of the true amplitude of variations in the source, rather than a convolution of true variability, observational uncertainties, and effects of finite sampling. Observational uncertainties and finite sampling will affect the accuracy of measurement of $\bar{m}$. Here, we decided to use the measured modulation index since the percentage difference between the modulation index ($\textit{m}$) and the intrinsic modulation index ($\bar{m}$) only ranged between 1$\%$--17$\%$. The modulation index for the target source varied between 5$\%$ and 18$\%$ and its uncertainty was taken from the uncertainty of the intrinsic modulation index. \par

\subsection{Reliability of the structure functions through Monte Carlo simulations}

\begin{figure*}
	\includegraphics[width=0.95\textwidth]{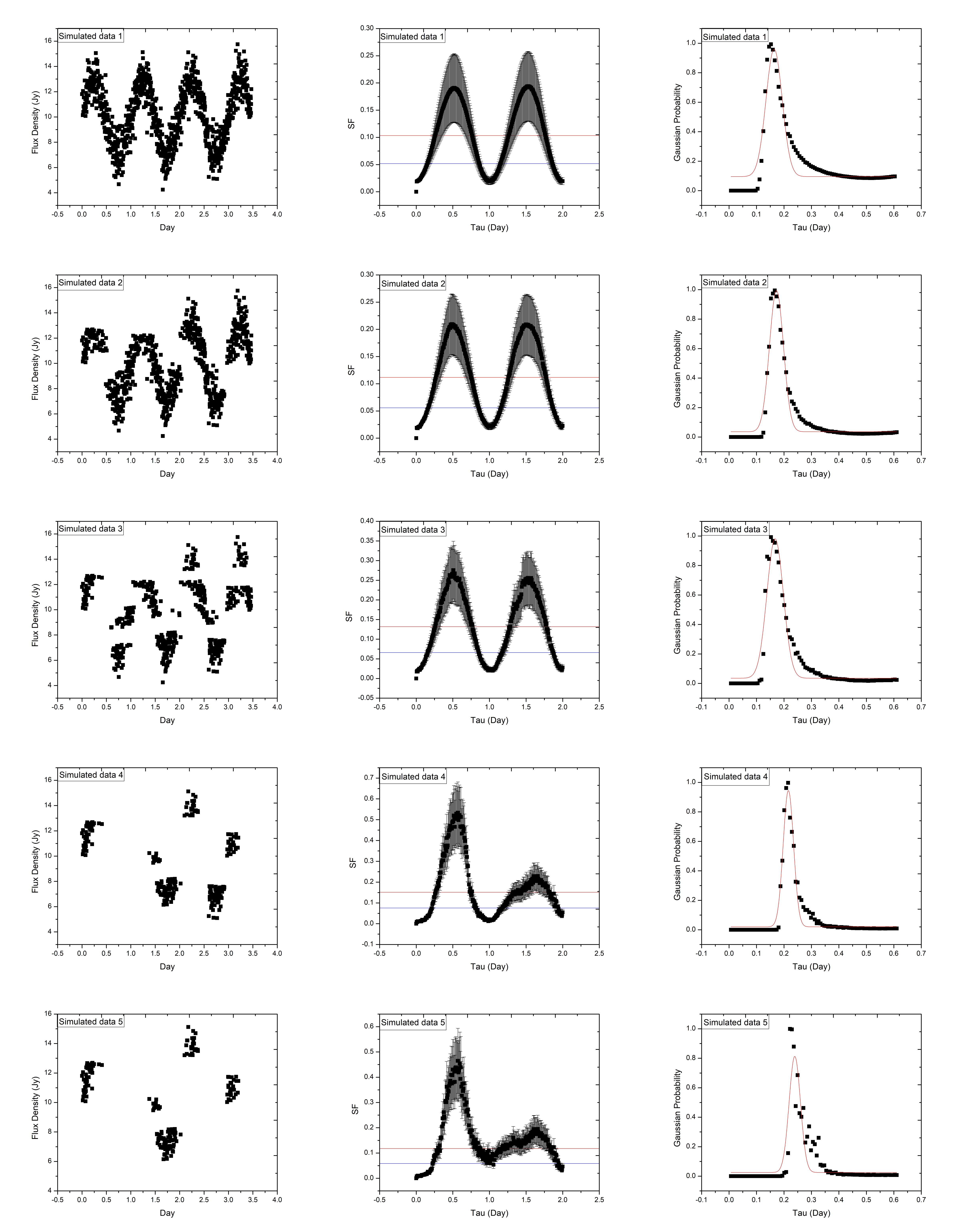}
	\centering
	\caption{Investigation of the reliability of Gaussian fitting to the Gaussian probability distribution of SF target source conducted using Monte Carlo simulations. 1000 iterations were performed using 5 datasets of light curves which contained 991, 772, 496, 234 and 178 data points (from top to bottom panels). The red and blue lines in the normalised SF plots represent the saturated ($2\sigma^{2}$) and variance values respectively. In this simulation, the initial model was a simple sinusoid with mean flux density of 10~Jy, a modulation amplitude of 3 Jy, a period of 0.16 days (equivalent to $T_{\textnormal{period}}$/4), a light-curve duration of 3.5 days and constant Gaussian white noise of amplitude 1 Jy.}
	\label{fig:monte_carlo_1}
\end{figure*}

An investigation of the reliability of Gaussian fitting to the Gaussian probability distribution of the target source SF was performed through Monte Carlo simulations. In this simulation, the initial model was a simple sinusoid with mean flux density of 10~Jy, a modulation amplitude of 3 Jy, a period of 0.16 days (equivalent to $T_{\textnormal{period}}$/4), a light-curve duration of 3.5 days and constant Gaussian white noise of amplitude 1 Jy. The light-curve duration of 3.5 days was chosen as it represents the period of first quarter of each observation block of the target source ($\sim$14 days) where the first ``scints'' can be observed. We adopted the first ``scints'' in estimating the variability timescale to reduce the uncertainties due to less sampling.  The average power of the signal ($P_{\textnormal{s}} = \textnormal{amplitude}^{2} / {2}$) and average power of uniform white noise ($P_{\textnormal{n}} = \textnormal{amplitude}^{2} / \sqrt{3}$) were 4.5 and 0.6 respectively, and the measured SNR (signal to noise ratio) of the signal is 8 ($ \textnormal{SNR} = 10.\log_{10} \left \{(P_{\textnormal{s}} ~-~ P_{\textnormal{n}}) / P_{\textnormal{n}} \right \} $) \citep{Sijbers-et-al-1996}. In order to examine how changes in data sampling frequency affect the light curves on the SF pattern, the peak of Gaussian fitting and the value of its FWHM (assumed as the uncertainty of the estimated characteristic timescale), 5 simulated datasets of light curves were created. These simulated light curves consisted of 991, 772, 496, 234 and 178 data points and had the same amplitude of Gaussian white noise and modulation period. The selected values represent the range of our observed data in each observation block of the target source. We then measured the SF for each of the simulated light curves and estimated the characteristic timescale as well as its uncertainty from the Gaussian fitting of the Gaussian probability distribution. Gaussian fitting of the Gaussian probability distribution was undertaken 1000 times through Monte Carlo simulations. Figure~\ref{fig:monte_carlo_1} shows 5 of these simulated light curves, the measured SF and the Gaussian fitting of the Gaussian probability distribution. We measured the peak of Gaussian fitting for each simulated light curves data to be 0.16, 0.17, 0.17, 0.22 and 0.24 days, for the light curves with 991, 772, 496, 234 and 178 data points, respectively. The uncertainties were 0.03 days for simulated light curves with 496 or more points and 0.02 for simulated light curves with 234 or 178 points. Our results suggest that a lack of data points located at the peaks and troughs of a light curve significantly changes the SF pattern, increasing the estimated characteristic timescale and decreasing its uncertainty by up to 30$\%$. Most of the light curves from the Ceduna dataset were similar in appearance to the 496 point simulated data sets; relatively well sampled but lacking some data at light curve peaks and troughs. This implies that the estimated characteristic timescale of Ceduna datasets are likely to be slightly overestimated compared to a fully sampled light curve.  \par

\section{Annual cycle in the variability timescales }

In this section, we define the annual modulation models, for isotropic and anisotropic scintillation patterns, we adopted in determining the annual cycle in the variability timescales. The annual modulation of radio IDV timescales can be explained as the relative motion of the Earth with respect to the scattering medium. This relative motion is comprised of 3 different velocities; the Earth's orbital motion around the Sun, the Sun's motion with respect to the local standard of rest (LSR) and the velocity of the scattering medium with respect to the LSR. The proper motion of the extragalactic sources is neglected ($\ll$ 1 km s$^{-1}$). If the peculiar velocity of the scattering medium is comparable to the Earth's orbital velocity ($\sim $ 30 km s$^{-1}$), then for part of the year, the velocity vectors are close to parallel, and six months later anti-parallel. When parallel, the apparent scattering medium velocity is low as seen from the Earth (longer characteristic timescale) and when anti-parallel the relative velocity is high (shorter characteristic timescale)  \citep{Jauncey-Macquart-2001}.  \par

\subsection{Isotropic model}  

Temporal variations in the flux density are caused by the relative transverse velocity of the scintillation pattern and the observer, $\textbf{\textit{V}}_{\textnormal{rel}}$. For an isotropic scattering medium, the characteristic timescale that is observed (as a function of day of year (DOY)) is determined by $\textbf{\textit{V}}_{\textnormal{rel}}$ and the characteristic spatial scale of the ISS pattern, $s_{\textnormal{o}}$:

\begin{equation}
	t_{\textnormal{char}}(DOY) = \dfrac{s_{\textnormal{o}}}{\textbf{\textit{V}}_{\textnormal{rel}}(DOY)}
	\label{eq:tchar isotropic}
\end{equation}

\noindent
where $\textbf{\textit{V}}_{\textnormal{rel}}$ has components in both right ascension, $\textbf{\textit{V}}_{\alpha}$, and declination, $\textbf{\textit{V}}_{\delta}$. This model function was defined using the approach of \citet{Carter2008}, who gives a thorough explanation of the geometries. In rectangular equatorial coordinates, a source with right ascension $\alpha$ and declination $\delta$ can be located with the position vector (for PKS\,B1144$-$379: right ascension ($\alpha$ = $11^{h}47^{m}01.4^{s}$) and declination ($\delta$ = $-38^{\circ} 12^{\arcmin} 11^{\arcsec}$)), $\hat{\textbf{\textit{e}}}_{\textnormal{source}}$, where:

\begin{equation}
\hat{\textbf{\textit{e}}}_{\textnormal{source}} = \left \{
\begin{tabular}{ccc}
$\cos$ $\delta$ $\cos$ $\alpha$ \\
$\cos$ $\delta$ $\sin$ $\alpha$ \\
$\sin$ $\delta$
\end{tabular}
\right \} 
\label{eq:position vector}
\end{equation}

The relative transverse velocity, $\textbf{\textit{V}}_{\textnormal{rel}}$, has components that are both parallel and perpendicular to the equatorial plane, $\textbf{\textit{V}}_{\alpha}$ and $\textbf{\textit{V}}_{\delta}$.  These components are in the directions $\hat{\textbf{\textit{e}}}_{\alpha}$ and $\hat{\textbf{\textit{e}}}_{\delta}$ respectively, where:

\begin{equation}
\hat{\textbf{\textit{e}}}_{\alpha} = \left \{
\begin{tabular}{ccc}
$-$$\sin$ $\alpha$ \\
$\cos$ $\alpha$ \\
0
\end{tabular}
\right \} \\
\hat{\textbf{\textit{e}}}_{\delta} = \left \{
\begin{tabular}{ccc}
$-$ $\cos$ $\alpha$ $\sin$ $\delta$ \\
$-$ $\sin$ $\delta$ $\sin$ $\delta$ \\
$\cos$ $\delta$
\end{tabular}
\right \}
\label{eq:position vector parallel and perpendicular}
\end{equation}

The relative velocity of the Earth and the scattering screen in the ISM, $\textbf{\textit{V}}_{\textnormal{Earth-ISM}}$ depends on the relative velocities of the Earth and the Sun, $\textbf{\textit{V}}_{\textnormal{Earth-Sun}}$ (a function of the DOY due to the orbital motion of the Earth), as well as the Sun and the LSR $\textbf{\textit{V}}_{\textnormal{Sun-LSR}}$:

\begin{equation}
\textbf{\textit{V}}_{\textnormal{Earth-ISM}}(DOY) = \textbf{\textit{V}}_{\textnormal{Earth-Sun}}(DOY) ~+~ \textbf{\textit{V}}_{\textnormal{Sun-LSR}} $-$ (\textbf{\textit{V}}_{\alpha}\hat{\textbf{\textit{e}}}_{\alpha} ~+~ \textbf{\textit{V}}_{\delta}\hat{\textbf{\textit{e}}}_{\delta})
\label{eq:velocity Earth-ISM}
\end{equation}

When projected onto the plane perpendicular to the line of sight, equation~\ref{eq:velocity Earth-ISM} gives the transverse relative velocity of the scintillation pattern:

\begin{equation}
\textbf{\textit{V}}_{\textnormal{rel}}(DOY) = \hat{\textbf{\textit{e}}}_{\textnormal{source}}~\times~(\textbf{\textit{V}}_{\textnormal{Earth-ISM}}(DOY)~\times~\hat{\textbf{\textit{e}}}_{\textnormal{source}})
\label{eq:transverse relative velocity of the scintillation pattern}
\end{equation}

\begin{figure}
	\centering
	\includegraphics[width=\columnwidth]{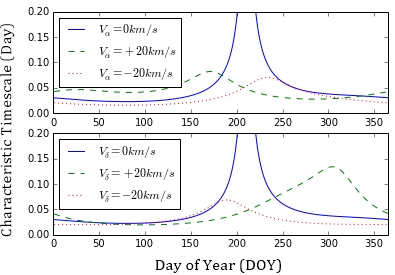}
	\caption{Dependence of the characteristic timescale as a function of day of year for PKS\,B1144$-$379 assuming an isotropic scattering screen with a variable velocity offset compared to the LSR. The top panel shows the predicted annual cycle fitting if the right ascension component of relative transverse velocity of the scintillation pattern and the observer ($\textbf{\textit{V}}_{\textnormal{rel}}$) is increased and decreased by 20 km s$^{-1}$ (with $\textbf{\textit{V}}_{\delta}$ = 0 km s$^{-1}$), while the bottom panel shows the effect of varying the declination component with the same values (with $\textbf{\textit{V}}_{\alpha}$ = 0 km s$^{-1}$). There is an apparent shift in the time of the year where the longest characteristic timescales are observed in each case.}
	\label{fig:isotropic_model}
\end{figure}

The predicted annual cycle depends on the coordinates of the source ($\hat{\textbf{\textit{e}}}_{\textnormal{source}}$). Figure~\ref{fig:isotropic_model} shows the predicted timescale as a function of DOY for PKS\,B1144$-$379, assuming an isotropic scattering screen with a variable velocity offset compared to the LSR. For a scattering screen moving with the LSR, the relative transverse velocity drops to almost zero near DOY 210, which result in a rapid increase in variability timescale (a near ``stand-still'' in the variations) around that time of the year. There are 3 fitted parameters obtained from the isotropic annual cycle model; the characteristic spatial scale of the phase variations on the scattering screen ($s_{\textnormal{o}}$) and both components of relative transverse velocity of the scintillation pattern and the observer ($\textbf{\textit{V}}_{\alpha}$ and $\textbf{\textit{V}}_{\delta}$). A non-linear least squares fit that assumed an isotropic (but non-stationary compared to the LSR) scattering screen was performed. We then implemented the Levenberg-Marquadt technique for non-linear least squares regression. Initially, each variability timescale data point was weighted with the inverse of the corresponding estimated uncertainty. However, as explained in \S 4, we also fit the data with uniform weighting for each data point. The $\chi^{2}$ values were calculated to quantify the accuracy of the model using the equations:

\begin{equation}
\chi^{2} = \sum_{\textnormal{N}}\dfrac{(f_{\textnormal{i}} ~-~ y_{\textnormal{i}})^{2}}{\sigma^{2}}
\label{eq:Chi-squared}
\end{equation}

\begin{equation}
\chi_{\textnormal{Red}}^{2} = \dfrac{\chi^{2}}{N ~-~ m}
\label{eq:Chi-squared Reduced}
\end{equation}

We take the square of the difference between the timescale estimate, $y_{\textnormal{i}}$, and the expected timescale according to the model, $f_{\textnormal{i}}$, scaled by the uncertainty in the measured timescale, $\sigma_{\textnormal{i}}$, for all $N$ timescale estimates. The reduced chi-squared value, $\chi_{\textnormal{Red}}^{2}$, takes into account the number of degrees of freedom (here \textit{m} is the number of free parameters in the model). For the isotropic scattering screen case, $\textit{m} = 3$.   \par

\subsection{Anisotropic model}  

\begin{figure}
	\centering
	\includegraphics[width=\columnwidth]{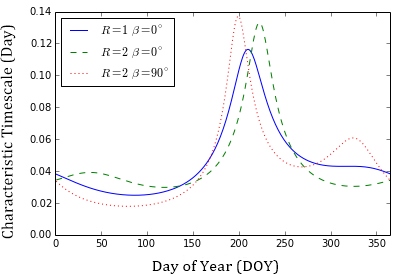}
	\caption{Dependence of characteristic timescale as a function of day of year for PKS\,B1144$-$379 assuming an anisotropic scattering pattern (with $\textbf{\textit{V}}_{\alpha}$ and $\textbf{\textit{V}}_{\delta}$ are equal to 5 km s$^{-1}$). The green dashed and red dotted lines represent the cases for an ellipse with an axial ratio of 2, where the major and minor axis respectively are aligned with the velocity of the scintillation pattern. For comparison, the isotropic scattering screen is shown in a blue solid line.}
	\label{fig:anisotropic_model}
\end{figure}

Next, an anisotropic model of the scattering screen in the ISM was considered. An anisotropic interference pattern is assumed to be in the form of ellipse with axial ratio \textit{R}, and its major axis inclined at angle $\beta$ to the direction of the relative transverse velocity \citep{Carter2008}. Figure~\ref{fig:anisotropic_model} shows the predicted timescale as a function of DOY for PKS\,B1144$-$379, assuming an anisotropic scattering screen with a variable velocity offset compared to the LSR. Similar to the isotropic model, the anisotropic model also predicts a sudden increase in the variability timescale around DOY of 200. The predicted timescale as a function of the DOY is:

\begin{equation}
t_{\textnormal{char}} (DOY) = \dfrac{s_{\textnormal{o}}\sqrt{R}}{\sqrt{\textbf{\textit{V}}_{\textnormal{rel}}^{2}(DOY) ~+~ (R^{2}-1)(\bar{\textbf{\textit{V}}}_{\textnormal{rel}}(DOY)~\times~ \hat{\textbf{\textit{S}}})^{2}}}
\label{eq:anisotropic model}
\end{equation}

\noindent
where $\hat{\textbf{\textit{S}}} = (\cos \beta, \sin \beta)$ and $\bar{\textbf{\textit{V}}}_{\textnormal{rel}} = (\textbf{\textit{V}}_{\parallel}, \textbf{\textit{V}}_{\perp})$. $\textbf{\textit{V}}_{\parallel}$ and $\textbf{\textit{V}}_{\perp}$ are the parallel and perpendicular components of $\textbf{\textit{V}}_{\textnormal{rel}}$ (relative transverse velocity of the scintillation pattern and the observer) to the equatorial plane. For the isotropic case where $\textit{R} = 1$ and $\beta = 0$, equation~\ref{eq:anisotropic model} becomes $t_{\textnormal{char}} = s_{\textnormal{o}} / \textbf{\textit{V}}_{\textnormal{rel}}$. There are 5 fitted parameters obtained from the anisotropic annual cycle model; the characteristic spatial scale of the phase variations on the scattering screen ($s_{\textnormal{o}}$), both components of relative transverse velocity of the scintillation pattern and the observer ($\textbf{\textit{V}}_{\alpha}$ and $\textbf{\textit{V}}_{\delta}$), the axial ratio ($\textit{R}$) and orientation angle ($\beta$). Due to two additional parameters $\textit{R}$ and $\textit{$\beta$}$ in the anisotropic model, the $\chi_{\textnormal{Red}}^{2}$ value is calculated with $\textit{m} = 5$. To estimate the uncertainty in the fitted parameters, we used the Levenberg-Marquadt algorithm, which involves estimation of the ($N \times m$) Jacobian matrix, $\textbf{\textit{J}}$, where:

\begin{equation}
\textbf{\textit{J}}_{\textnormal{ij}} = \dfrac{\partial f(x_{\textnormal{i}},\textbf{\textit{p}})}{\partial p_{\textnormal{j}}}
\label{eq:Jacobian matrix}
\end{equation}

\noindent
and $\textbf{\textit{p}}$ is a vector of length \textit{m} containing each parameter estimate. For this method, the uncertainty in the final estimates is defined as:

\begin{equation}
\sigma_{\textbf{p}} = \sqrt{\textnormal{trace}(\chi_{\textnormal{Red}}^{2} [\textbf{\textit{J}}^{\textbf{\textit{T}}} \textbf{\textit{W}} \textbf{\textit{J}}]^{-1})}
\label{eq:uncertainty fitted parameters}
\end{equation}

\noindent
where $\textit{W}$ is a diagonal matrix consisting of the weights of each data point, which were initially set to the inverse magnitude of the corresponding uncertainties, and later set to be uniform (unweighted fitting, as discussed in \S 4).   \par

\subsection{Reliability of the annual cycle fitting}

\begin{figure*}
	\includegraphics[width=1.0\textwidth]{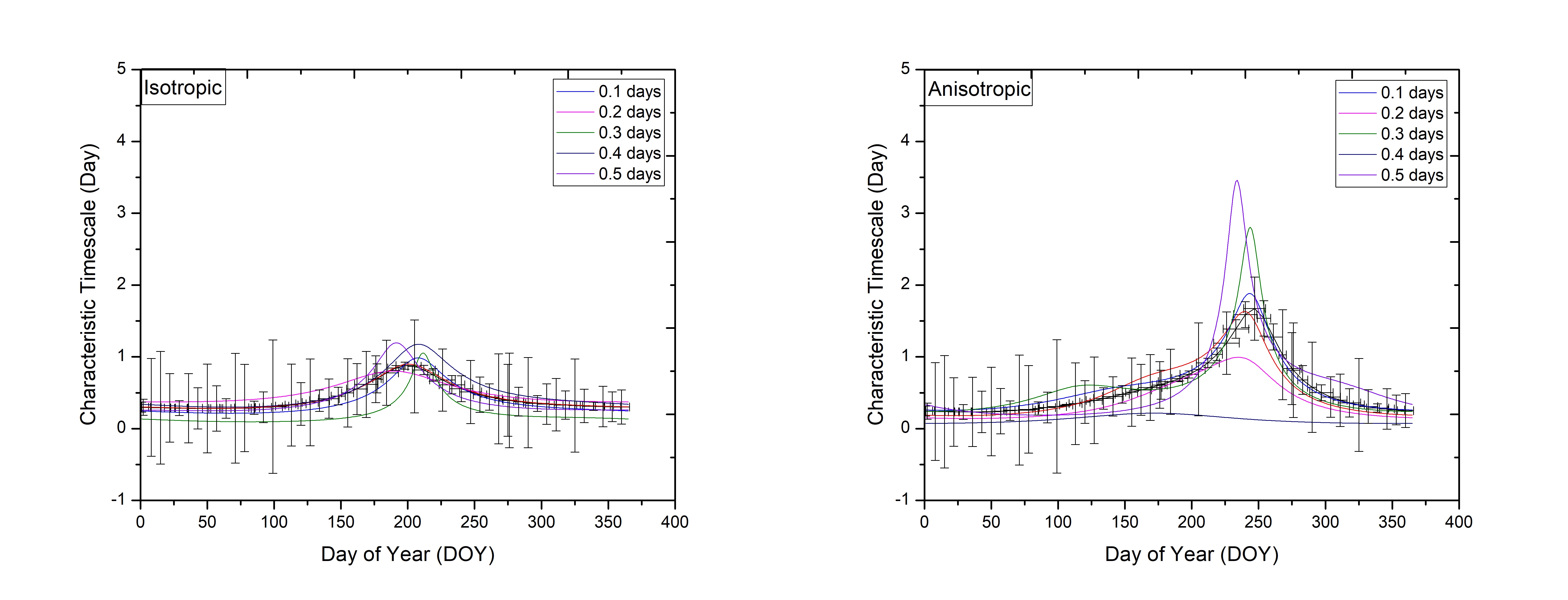}
	\centering
\caption{Investigation of the effect of data points distribution on isotropic (left panel) and anisotropic annual cycle fitting (right panel) conducted using 1000 iterations of Monte Carlo simulations. The simulated characteristic timescales were randomly varied by a constant amount for each simulation.  The amplitude of the variations was between $\pm$ 0.1 and 0.5 day, in 0.1 day steps. The estimate of the slowest period in the isotropic annual cycle varied between DOY 191--212 and the ``true'' DOY for the longest timescale of simulated isotropic model was DOY of 200. Whereas the estimate of the slowest period in the anisotropic annual cycle varied between DOY 234--244, compared to the ``true'' one which was DOY of 248.}
	\label{fig:monte_carlo_2}
\end{figure*}

\begin{figure*}
	\includegraphics[width=\textwidth]{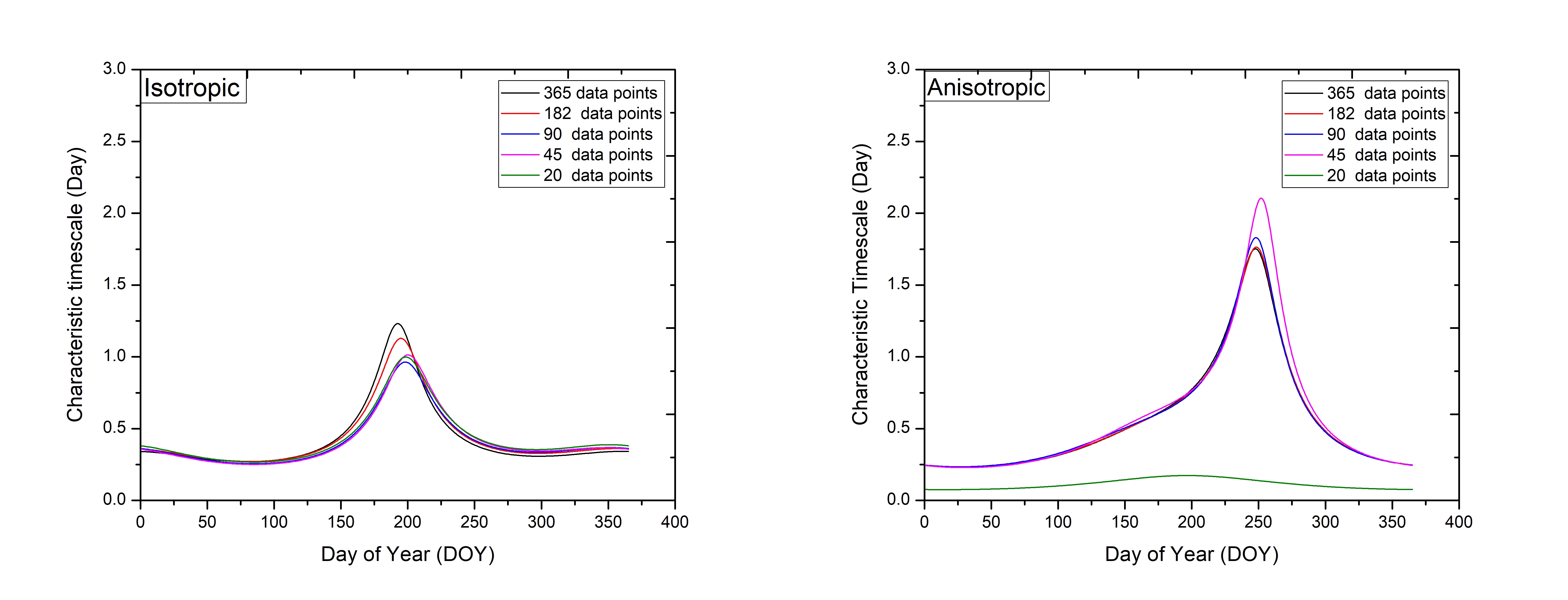}
	\centering
\caption{Investigation of the effect of the number of data points on the accuracy of the isotropic and anisotropic annual cycle fitting conducted using 1000 iterations of Monte Carlo simulations. Five datasets containing 365, 182, 90, 45 and 20 characteristic timescale estimates with constant values of uncertainties were created. For the isotropic model, the annual cycle fits were very similar and the estimate of the slowest period of the annual cycle varied slightly from DOY 192 to 201 as the number of timescale estimates was decreased. As for the anisotropic model, it showed much greater sensitivity, with the annual cycle fittings changed remarkably with estimate of the slowest period of the annual cycle varying in the range DOY 200--252. The ``true'' DOY for the longest timescale of these simulated isotropic and anisotropic models were DOY of 193 and 248, respectively.}
	\label{fig:monte_carlo_3}
\end{figure*}

Several tests to examine the reliability of the annual cycle fitting for both isotropic and anisotropic models were also performed using Monte Carlo simulations. The first test was to examine the effect of the distribution of data points on the annual cycle fitting (variations of the characteristic timescales). In this simulation, we sampled an annual cycle curve at 55, evenly spaced, points over the year.  The characteristic timescales ranged from 0.01--1.60 days.  The longest timescales  were around DOY of 200 and 248 for isotropic and anisotropic fitting models. We then added noise to the sampled characteristic timescales.  For each simulation the amplitude of the random noise was fixed at a value which was varied between $\pm$0.1 and 0.5 day in steps of 0.1 days. Average annual cycle fittings were determined through 1000 iterations of Monte Carlo simulations for each noise amplitude. Figure~\ref{fig:monte_carlo_2} shows the average annual cycle fittings for both isotropic and anisotropic models. We found that for both isotropic and anisotropic annual cycle models, the estimate of the slowest period in the annual cycle was in error by 6$\%$. However, the anisotropic annual cycle model, with more free parameters, was much more susceptible to the noise in the characteristic timescale estimates for random noise more than 0.5 days. These results indicate that the uncertainty in the characteristic timescale estimates plays a major role in the accuracy of the annual cycle fitting. We also investigated how the accuracy of the annual cycle fitting depended on the number of characteristic timescale measurements. Five sets of data of the characteristic timescale measurements with constant values of uncertainties were created (Fig.~\ref{fig:monte_carlo_3}). We then determined the average annual cycle fitting for each of the dataset through 1000 iterations of Monte Carlo simulations. Figure~\ref{fig:monte_carlo_3} illustrates the average annual cycle fitting for both isotropic and anisotropic models for all five simulated datasets. For the isotropic model, the annual cycle fits were very similar and the estimate of the slowest period of the annual cycle was only 4$\%$ off as the number of timescale estimates was decreased. On the other hand, the anisotropic model showed much greater sensitivity, with the annual cycle fittings changing remarkably. The estimated slowest period of the annual cycle had a 19$\%$ error as we reduced the timescale values. Our results suggest that in order to accurately fit the annual cycle, at least 20 independent measurements of characteristic timescale covering the slow-down period are required. Estimation of fitted parameters from the annual cycle fitting can be improved by more continuous monitoring.  \par

\section{Results}

\begin{table*}
	\caption{Characteristic timescales ($t_{\textnormal{char}}$ = $T_{\textnormal{period}}$/4) of PKS\,B1144$-$379 for year 2003 (the full table of characteristic timescale for the period 2003--2011 is available online). Col. 1 gives the observation year,  Col. 2 the day of year at the mid-point of the observing session, Col 3 is the duration of the session, divided by two, which we use to plot the time range used for each timescale estimate, Col. 4 and 5 the measured characteristic timescale and its uncertainty, Col. 6 and 7 the observed mean flux density and its uncertainty, Col. 8 and 9 the modulation index and its uncertainty. For each period of observations of length $T_{\textnormal{obs}}$, the uncertainty of observations (as a function of day of year) is given by the median date ($\delta$DOY = $T_{\textnormal{obs}}$/2). The measurements of the characteristic timescales ($t_{\textnormal{char}}$), modulation index (\textit{m}) and its uncertainties were defined in Section 2.2.}
	\centering
	\label{table:characteristic_timescales}
	\begin{tabular}{rrrrrrrrr}
		\hline
		Year & DOY & $\delta$ DOY & $t_{\textnormal{char}}$ (day) & $\delta$$t_{\textnormal{char}}$ (day) & $\bar{S}$ (Jy) & $\delta$$\bar{S}$ (Jy)  & $\textit{m}$ ($\%$) &  $\delta$$\textit{m}$ ($\%$)   \\
		\hline
		2003 &         99 &          5 &      1.30 &      0.35 &      2.05 &      0.16 &      7.60   &   0.70  \\
		
		&        124 &         10 &      1.00 &      0.22 &      2.03 &      0.22 &      10.90   &  1.00   \\
		
		&        144 &          3 &      0.50 &      0.03 &      2.06 &      0.12 &      5.70   &  0.20  \\
		
		&        164 &          5 &       0.54 &      0.10 &      1.99 &       0.16 &       8.00   &  0.70   \\
		
		&        198 &          5 &      0.47 &      0.10 &      2.10 &      0.17 &      7.80   &  0.70   \\

		\hline
	\end{tabular}
\end{table*}
In this section, we will discuss the results of annual cycle fitting and the source-intrinsic variability of PKS\,B1144$-$379.

\subsection{Annual cycle fitting}

In Table~\ref{table:characteristic_timescales}, we summarize results of the observations of PKS\,B1144$-$379 taken as part of the COSMIC observing project between 2003 to 2011. The mean flux density of PKS\,B1144$-$379 at 6.7 GHz over the period 2003--2011 was found to vary by a factor of 4. The long-term monitoring of the flux density of this source shows that there were two flares, which commenced in 2005 November and 2008 August. Further investigation of the long term variability and high resolution imaging is part of a future publication. The characteristic timescale of the source (equivalent to $T_{\mbox{period}}/4$) was measured to vary from 0.22--3.37 day, with the longest timescale occurring during the first flare (2005 November). We found the modulation index of PKS\,B1144$-$379 ranged from 5--18 $\%$. \par

\begin{figure*}
	\centering
	\includegraphics[width=\textwidth]{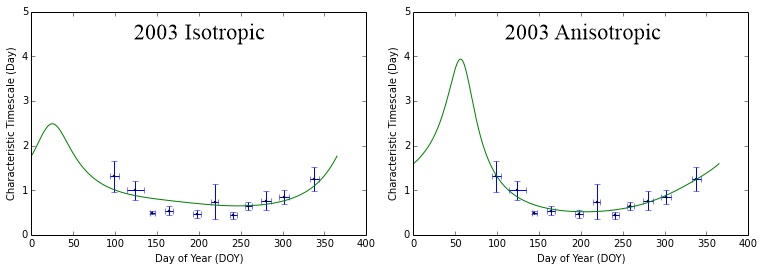}
	\caption{Annual cycle fits of isotropic and anisotropic models for PKS\,B1144$-$379 for year 2003 (the remaining panels for the period 2003--2011 are available online). Plotted in green is the unweighted annual cycle fits. Due to the limited data in 2011, only an isotropic annual cycle model can be fitted. Data taken from the COSMIC observing project.}
	\label{fig:annual_cycle}
\end{figure*}

\begin{figure*}
	\centering
	\includegraphics[width=\textwidth]{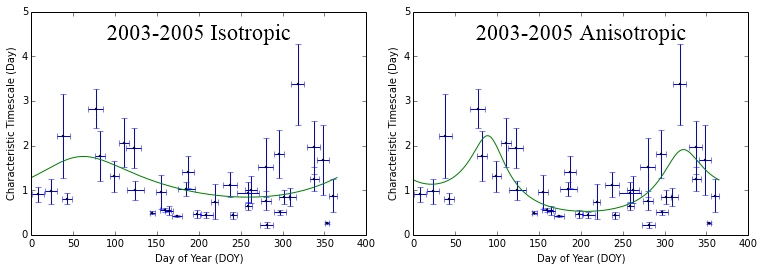}
	\caption{Annual cycle fits of isotropic and anisotropic models for PKS\,B1144$-$379. Plotted in green is the unweighted annual cycle fittings. Data points from 3 years observations were combined and plotted in order to verify the detection of the annual cycle (the remaining panels for the period 2003--2011 are available online). Data taken from the COSMIC observing project.}
	\label{fig:annual_cycle_combine}
\end{figure*}

Figure~\ref{fig:annual_cycle} shows the variability timescale versus day of the year of PKS\,B1144$-$379 from 2003--2011 fitted with both isotropic and anisotropic annual cycle models. We also performed a comparison between weighted and unweighted fitting. The annual cycle in the variability timescale was found to be more prominent when the fitting does not take into account the estimated uncertainties. The estimated uncertainty in the timescale depends on the duration of the observing session compared to the characteristic timescale.  The majority of our observing sessions were 10--15 days in duration, so the result is that the sessions corresponding to longer characteristic timescales (slow-down periods) have larger estimated uncertainties and hence have lesser weight in the weighting fitting process (i.e. the fit is down-weighting the slow-down period). Our Monte Carlo simulation shows that the slow-down periods are most important to constrain the parameters of the annual cycle as the absence of these crucial data points increases the fitted value for the spatial scale of the screen by a factor of 6, inflating the $\textbf{\textit{V}}_{\alpha}$, $\textbf{\textit{V}}_{\delta}$ and $\textit{R}$ to unrealistic values and decreases the $\beta$ by a factor of 2. Therefore, in this paper, we chose to determine the fitted parameters from the unweighted annual cycle fitting. Although for some years there are fewer data points than for others, the plots show consistent peaks of longer characteristic timescales (slow scintillation) around DOY 50--100 and 300--350. Shorter timescales (fast scintillation) are observed between DOY 150--250. These patterns are seen also by combining the data in 3 year groupings and combining the entire 9 years of observations (Fig.~\ref{fig:annual_cycle_combine}). The persistent detection of an annual modulation pattern is a strong evidence that interstellar scintillation is the cause of the intraday variability observed in PKS\,B1144$-$379. \par

The observed pattern in the timescale for PKS\,B1144$-$379 is quite different to those shown in Figures~\ref{fig:isotropic_model} and \ref{fig:anisotropic_model}, for isotropic and anisotropic scintillation models with screens moving at velocities close to the local standard of rest (\S 3). Those models predict ``slow scintillation'' to be observed around DOY 210. In contrast, the observed variability timescales were found to be shortest during this particular period.  We have examined the reduced $\chi^{2}$ values for both the isotropic and anisotropic models for each of the years.  The reduced $\chi^{2}$ values for the isotropic fit vary by a factor of 4 (excluding 2009 and 2011, where we have insufficient sampling of the change in characteristic timescale over the year to obtain a meaningful fit). For the anisotropic fit the reduced $\chi^{2}$ varies by a factor of 6.  For most years both the isotropic and anisotropic fits have reduced $\chi^{2}$ greater than 2, demonstrating that the model does not fit the data particularly well.  The most likely reason for the relatively poor fit of the annual cycle models is that the source structure evolves over the year as indicated by changes in both source flux density and structure index \citep{Shabala-et-al-2014}.  Further evidence for this is that the highest reduced $\chi^{2}$ values are seen for 2005 and 2008, the two years corresponding to the outbursts.  It should be noted that the longest timescale observed over the 8 years covered in the current work coincides with the start of the 2005 November flare. At this point, the modulation index is large in addition to the timescale being long. This particular one period might be just an anomalously large "scintle" (large and slow modulation) due to the stochastic nature of the scattering. Another possibility is that the scintillating component has both significantly brightened (compact fraction is large, hence large modulation) and expanded (longer timescale). \par

\begin{table*}
	\caption{Fitted parameters of PKS\,B1144$-$379 for both unweighted isotropic and anisotropic annual cycle models (the full table of fitted parameters for the period 2003--2011 is available online). Col. 1 gives the label of the annual cycle model, Col. 2 and 3 the observing year and length scale of scintillation pattern ($s_{\textnormal{o}}$), Col. 4 and 5 the RA ($\textbf{\textit{V}}_{\alpha}$) and Dec ($\textbf{\textit{V}}_{\delta}$) components of the relative transverse velocity of the scattering screen and the observer, Col. 6 and 7 the axial ratio, $\textit{R}$ and orientation angle $\beta$, and Col. 8 and 9 the position angle of the anisotropy scattering screen and reduced $\chi^{2}$. For both cases of isotropic and anisotropic annual cycle models, it implies that the scattering screen in the ISM has a high velocity offset compared to LSR. This is consistent with our expectations as the observed characteristic timescales are contrast with the predicted models which contribute to high values of reduced $\chi^{2}$. For some of the observing years, $\textit{R}$ has gone to zero (or infinity) and both components of the relative transverse velocity of the scintillation pattern and the observer ($\textbf{\textit{V}}_{\alpha}$ and $\textbf{\textit{V}}_{\delta}$) blows out to unrealistic values. The fitted parameters are poorly constrained due to the evolution of the target source. Date taken from COSMIC project observed from 2003 through 2011.}
	\centering
	\label{table:fitted_parameters}
		\begin{tabular}{ccccccccc}
			\hline
			Model &  Year &  $s_{\textnormal{o}}$  & $\textbf{\textit{V}}_{\alpha}$  &  $\textbf{\textit{V}}_{\delta}$  &  $\textit{R}$ &  $\beta$  & PA  &  $\chi^{2}$  \\
			
			&  & $(10^{6}$ $kms^{-1})$  &  (km$s^{-1}$)  &  (km$s^{-1}$)  & & (rad)  &  (deg) &  \\
			\hline
			
			Isotropic &       2003 & 3.10$\pm$2.22 & 43.50$\pm$28.12 & 27.26$\pm$25.31  &            &            &            &      1.84 \\
			
			Anisotropic &            & 3.00$\pm$4.80 & 56.74$\pm$126.98 &  $-$2.11$\pm$91.85 &  0.26$\pm$0.69 & 0.97$\pm$0.32  &      34.40 &       1.14 \\
			
			&            &            &            &            &            &            &            &            \\
			
			Isotropic &       2004 & 1.96$\pm$0.42 & 35.94$\pm$2.84   &  $-$2.42$\pm$3.09 &            &            &            &      4.91 \\
			
			Anisotropic &            & 2.69$\pm$1.10 & 38.53$\pm$32.41 &  $-$4.33$\pm$25.66 & 0.25$\pm$0.27 & 0.91$\pm$0.14 &      37.93 &      4.84 \\
			
			&            &            &            &            &            &            &            &            \\
			
			Isotropic &       2005 & 13.55$\pm$21.44 & 116.88$\pm$192.73 & 76.39$\pm$138.50 &            &            &            &      7.54 \\
			
			Anisotropic &            & 28.10$\pm$1276.00 & 2204.84$\pm$199330.83 &  $-$1980.06$\pm$181981.62 &  0.001$\pm$0.10 & 0.83$\pm$0.06 &      42.37 &      5.76 \\
			
			&            &            &            &            &            &            &            &            \\

			\hline 
		\end{tabular}
\end{table*}

In Table~\ref{table:fitted_parameters}, we summarize results of the fitted parameters for both isotropic and anisotropic annual cycle models for PKS\,B1144$-$379. We list the fitted parameters for each year from 2003 through 2011, inclusive. As previously mentioned, due to the larger formal uncertainty when the source exhibits longer characteristic timescales, we chose to determine the fitted parameters from the unweighted annual cycle fitting. For the year 2011, the anisotropic model could not be fitted due to insufficient data. For some years, the fitted parameters are not at all well constrained (larger by a factor of $>>$ 100), whereas for other years (2006 and 2007) they appear to be reasonably well constrained (by a factor of 3). The large range of fitted parameters is due to some unrepresentative outliers and small amount of data points. We also found that for some observing years (2005 and 2010), the screen velocity has unrealistic values and the $\textit{R}$ has gone to zero (or infinity), demonstrating that the fitted parameters were not well constrained, likely due to intrinsic source evolution which is entangled with the interstellar scintillation. Both these periods coincide with new flares which are likely to correspond to ejection of new plasma components. The emergence of new components during the source evolution leads to lengthening of the characteristic timescales. This will cause ambiguity in the fitted parameters as the shape of the slow down period is critical in determining the ISM velocity and anisotropy in the scintillation pattern \citep{Bignall-et-al-2003}. The lower reduced $\chi^2$ for the anisotropic models shows that it is able to provide a better fit to the data.  \par

\begin{figure*}
	\centering
	\includegraphics[width=\textwidth]{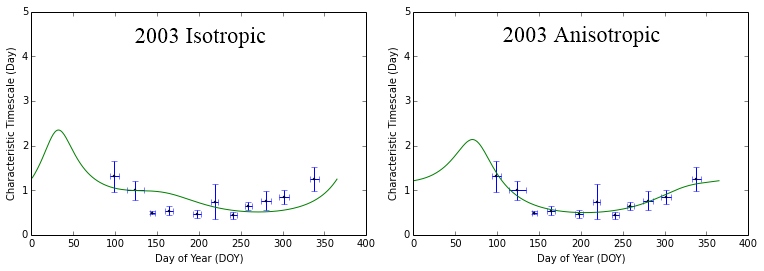}
	\caption{Unweighted annual cycle fittings of isotropic and anisotropic models for PKS\,B1144$-$379 for year 2003 (the remaining panels for the period 2003--2011 are available online). The $\textbf{\textit{V}}_{\alpha}$ and $\textbf{\textit{V}}_{\delta}$ were fixed to be 30 and 7.7 km$s^{-1}$ respectively based on 2006--2007 data. Data taken from the COSMIC observing project.}
	\label{fig:annual_cycle_constrained}
\end{figure*}

\begin{figure*}
	\centering
	\includegraphics[width=\textwidth]{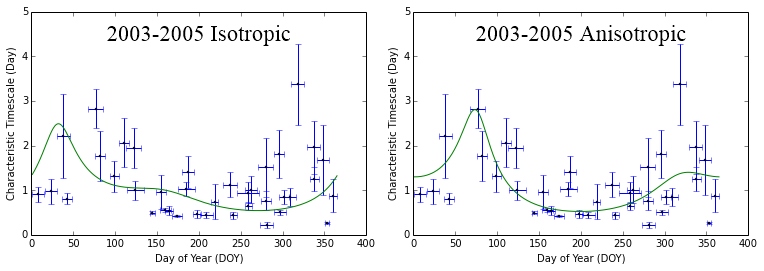}
	\caption{Unweighted annual cycle fitting of isotropic and anisotropic models for PKS\,B1144$-$379. Data points from 3 years observations were combined and plotted in order to verify the detection of the annual cycle (the remaining panels for the period 2003--2011 are available online.). The $\textbf{\textit{V}}_{\alpha}$ and $\textbf{\textit{V}}_{\delta}$ were fixed to be 30 and 7.7 km$s^{-1}$ respectively based on 2006--2007 data. Data taken from the COSMIC observing project.}
	\label{fig:annual_cycle_combine_constrained}
\end{figure*}

We investigated the effect of fixing some of the model parameters, as for some of the years the fitted values for the screen velocity and the axial ratio $\textit{R}$ of the scattering screen produce unrealistic values. We chose to fix $\textbf{\textit{V}}_{\alpha}$ and $\textbf{\textit{V}}_{\delta}$ to be 30 and 7.7 km s$^{-1}$, respectively. These values come from the fitted parameters of observing year 2006 and 2007 where they appear to be reasonably well constrained. A persistent slow-down period near DOY 50 and towards the end of the year (Fig.~\ref{fig:annual_cycle} and Fig.~\ref{fig:annual_cycle_combine}) is a strong indication that there is a repeating annual cycle in the variability timescale of PKS\,B1144$-$379. Therefore, it seems reasonable to assume that the velocity and other properties of the scattering screen remain unchanged over the entire observing period. Figure~\ref{fig:annual_cycle_constrained} shows the variability timescales versus day of the year for PKS\,B1144$-$379 from 2003 through 2011 fitted with fix values of $\textbf{\textit{V}}_{\alpha}$ (30 km$s^{-1}$) and $\textbf{\textit{V}}_{\delta}$ (7.7 km$s^{-1}$). Figure~\ref{fig:annual_cycle_combine_constrained} displays plots of the variability timescale combining data in three year groups and for the entire 9 years of observations. Both Figure~\ref{fig:annual_cycle_constrained} and Figure~\ref{fig:annual_cycle_combine_constrained} show the unweighted annual cycle fittings. Overall, we observed an improvement in the annual cycle fitting as the optimum fittings now pass through almost all the data points, especially at the peak of ``slow-scintillation''. The fitted parameters values were also found to be better constrained. In Table~\ref{table:fitted_parameters_constrained}, we summarize results of the fitted parameters for both unweighted isotropic and anisotropic annual cycle models of PKS\,B1144$-$379, with the screen velocity fixed. Overall, the fitted parameters appear to be reasonably well constrained. The 2011 data can now be fitted with the anisotropic annual cycle model as we only  have 3 parameters that are allowed to freely vary ($s_{\textnormal{o}}$, $\textit{R}$ and $\beta$). \par

\begin{table*}  
	\caption{Fitted parameters of PKS\,B1144$-$379 for both isotropic and anisotropic annual cycle models (the full table of fitted parameters for the period 2003--2011 is available online). The fitted parameters were well constrained when we fixed the values of both velocities component of scattering screen $\textbf{\textit{V}}_{\alpha}$ and $\textbf{\textit{V}}_{\delta}$ to be 30 and 7.7 km$s^{-1}$ respectively based on 2006-2007 data. Data taken from the COSMIC project observed from 2003 through 2011.}
	\centering
	\label{table:fitted_parameters_constrained}
		\begin{tabular}{ccccccccc}
			\hline
			Model &  Year &  $s_{\textnormal{o}}$  & $\textbf{\textit{V}}_{\alpha}$  &  $\textbf{\textit{V}}_{\delta}$  &  $\textit{R}$ &  $\beta$  & PA  &  $\chi^{2}$  \\
			
			&  & $(10^{6}$ $kms^{-1})$  &  (km$s^{-1}$)  &  (km$s^{-1}$)  & & (rad)  &  (deg) &  \\
			\hline
			
			Isotropic &       2003 & 1.84$\pm$0.22 &         30 &        7.7 &            &            &            & 5.80           \\
			
			Anisotropic &            & 1.86$\pm$0.09 &         30 &        7.7 & 0.36$\pm$ 0.06 & 0.97$\pm$0.05 &      34.42 &       0.94 \\
			
			&            &            &            &            &            &            &            &            \\
			
			Isotropic &       2004 & 2.12$\pm$0.45 &         30 &        7.7 &            &            &            &         8.43   \\
			
			Anisotropic &            & 2.22$\pm$0.33 &         30 &        7.7 & 0.25$\pm$0.12 & 0.89$\pm$0.12 &         30 &       7.49 \\
			
			&            &            &            &            &            &            &            &            \\
			
			Isotropic &       2005 & 1.88$\pm$0.42 &         30 &        7.7 &            &            &            &         9.68   \\
			
			Anisotropic &            & 2.43$\pm$0.48 &         30 &        7.7 & 0.27$\pm$0.13 & 1.07$\pm$0.16 &      28.69 &      8.43 \\
			
			&            &            &            &            &            &            &            &            \\
			
			\hline 
		\end{tabular}
\end{table*}

\subsection{Source-intrinsic variability}

\begin{figure}
	\centering
	\includegraphics[width=\columnwidth]{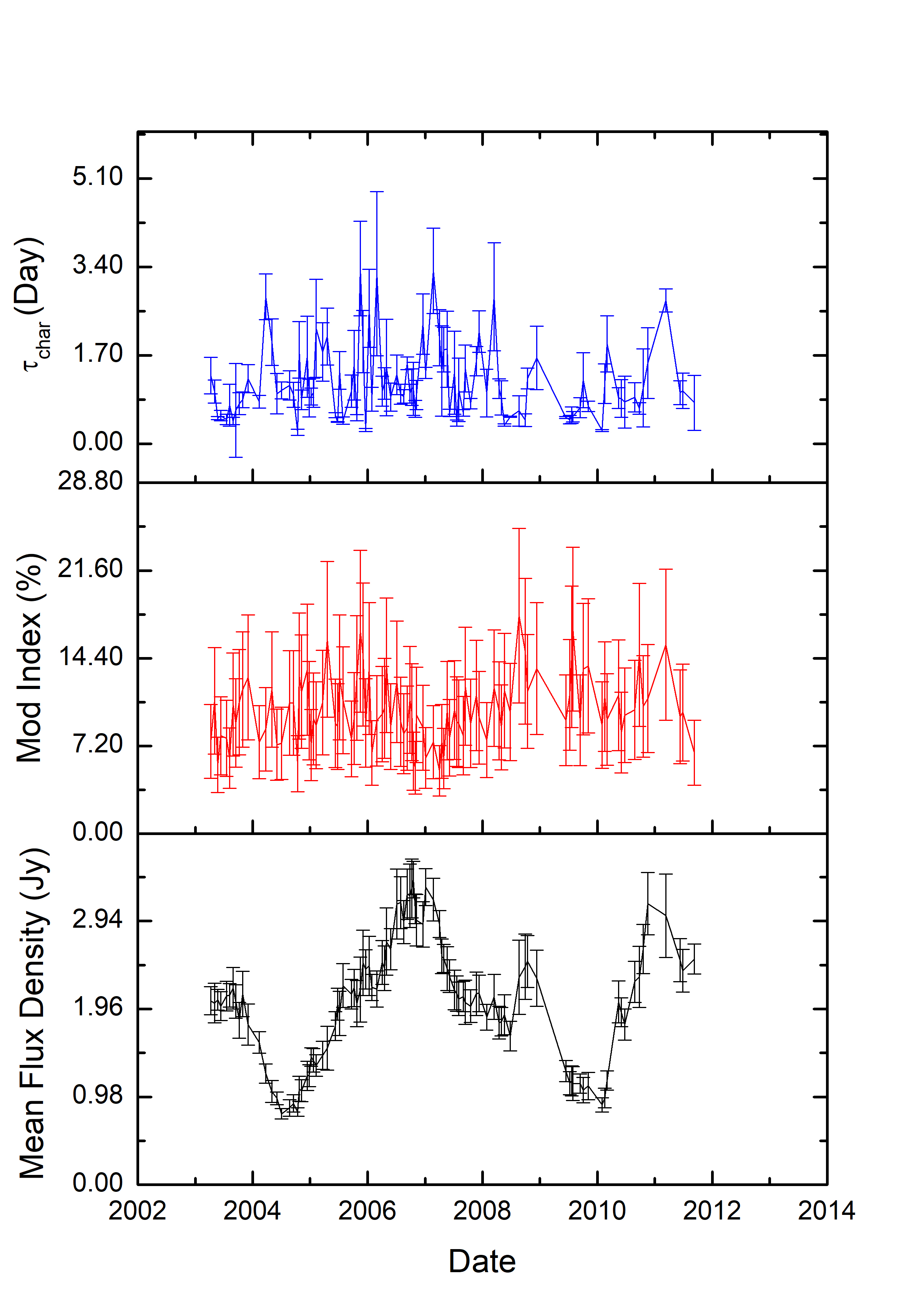}
	\caption{Long-term monitoring of variations in the flux density of PKS\,B1144$-$379 between 2003 and 2011. Each panel illustrates (top to bottom); characteristic timescale ($t_{\textnormal{char}}$), modulation index ($\%$) and mean flux density.}
	\label{fig:long_term_variations}
\end{figure}

\begin{figure}
	\centering
	\includegraphics[width=\columnwidth]{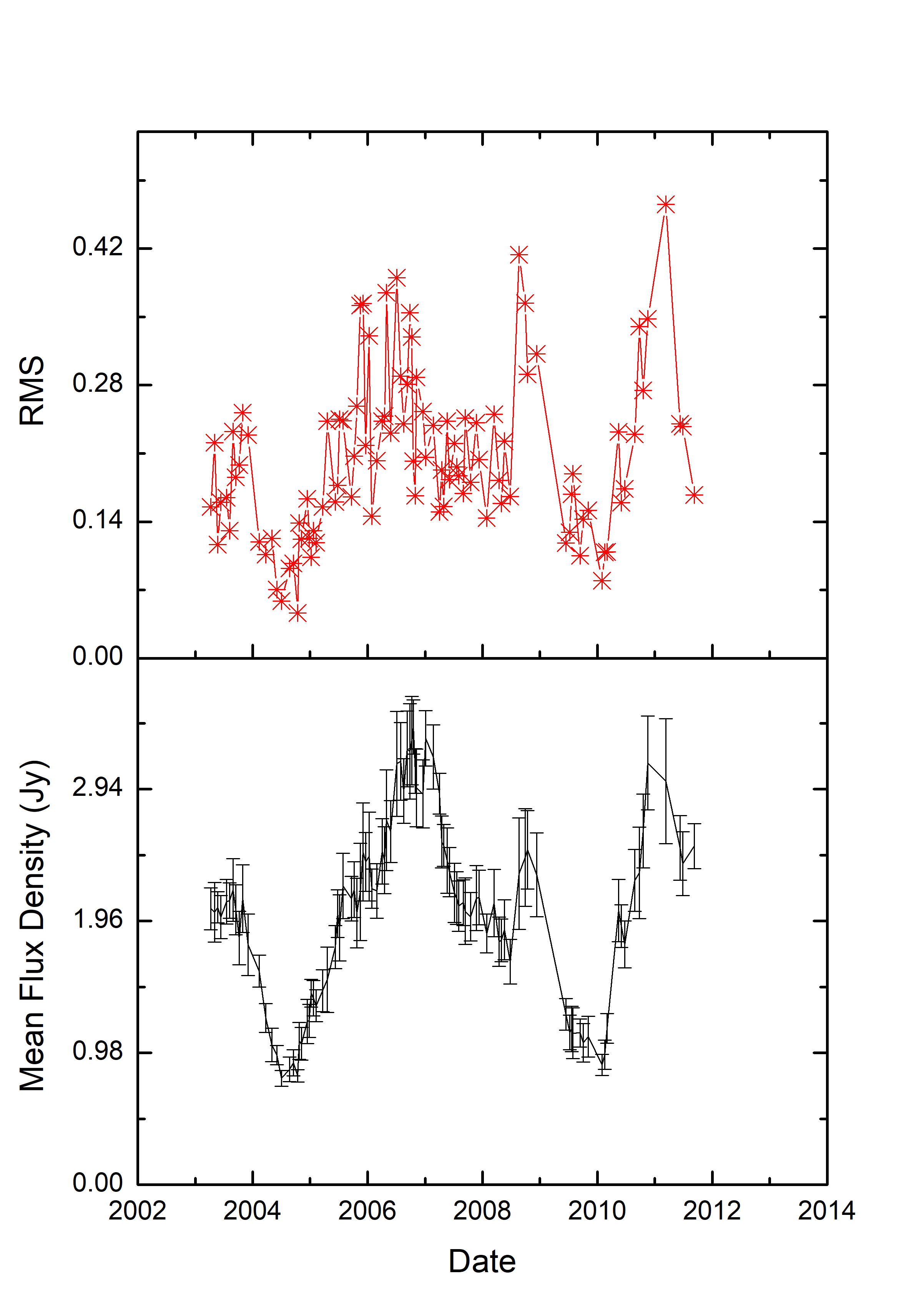}
	\caption{Long-term monitoring of variations of flux density for PKS\,B1144$-$379 between 2003 and 2011. Top and bottom panels show the RMS and mean flux density respectively.}
	\label{fig:flux_density_rms}
\end{figure}

\begin{figure}
	\centering
	\includegraphics[width=\columnwidth]{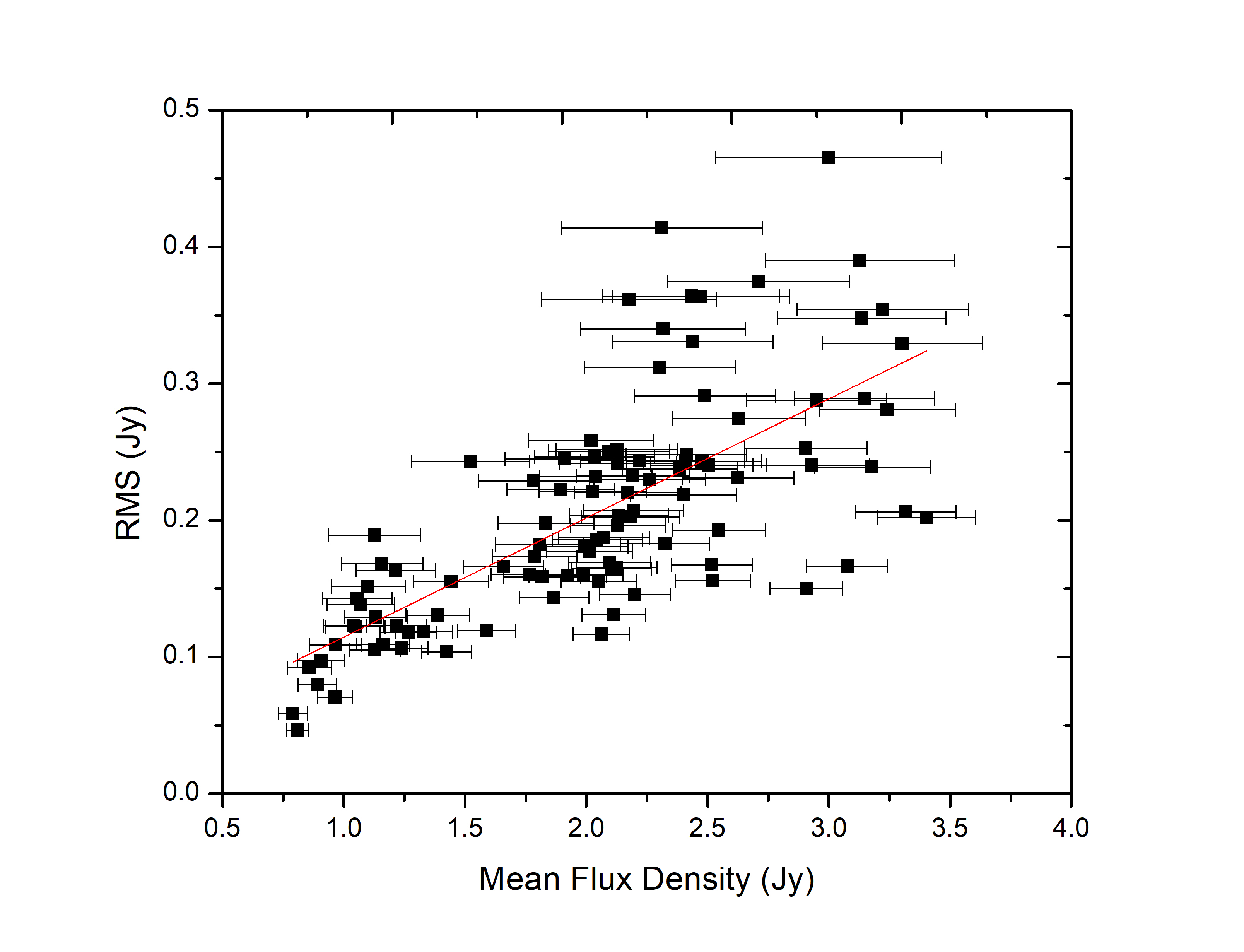}
	\caption{Linear correlation between the RMS and the mean flux density of PKS\,B1144$-$379. The correlation coefficient between these two variables was measured to be 0.7.}
	\label{fig:correlation_RMS_mean_flux_density}
\end{figure}

To investigate whether there are correlations between the radio IDV and long-term variations in the total flux density, we plotted the temporal changes in the mean flux density, modulation index and characteristic timescales ($t_{\textnormal{char}}$) against time (Fig.~\ref{fig:long_term_variations}). No correlation was found between the mean flux density and the modulation index as well as the mean flux density and the characteristic timescale. Figure~\ref{fig:flux_density_rms} shows the RMS of the flux density and the mean flux density on the same temporal scale and a clear correlation with correlation coefficient of 0.7 between the two is evident (refer to Figure~\ref{fig:correlation_RMS_mean_flux_density}). This demonstrates that the amplitude of the radio IDV phenomenon is larger when the mean flux density is high and hence that the radio IDV is entangled with the temporal evolution of the source. Higher values of the modulation index imply that the source becomes more compact and scintillates more, or that the fraction of the flux density in the compact component increases. Our results are in good agreement with \citet{Carter-at-al-2009}. They reported a loose correlation between the mean flux density and modulation index for PKS\,B1519$-$263 which led them to suggest a significant fraction of the change in the source mean flux density was due to evolution of the compact scintillating core. \citet{Bignall-et-al-2003} reported a large increase in the RMS variations is expected if the increase in the total flux density is due to an increase in the flux density of the scintillating component. In addition, if the scintillating component expands, this will manifest as an increase in the characteristic timescale, which can also be detected \citep{Liu-et-al-2012}.  \par

The long-term variations in the mean flux density of PKS\,B1144$-$379 are most likely due to intrinsic source evolution. The amplitude of the short timescale variability due to ISS depends on the fraction of the source flux density in the scintillating components, the angular size of the source, the observed frequency, and the ratio of the observed frequency to the transition frequency between weak and strong scattering \citep{Walker1998}. PKS\,B1144$-$379 is located at Galactic coordinates of $289.24^{o}$ (longitude), $22.95^{o}$ (latitude) and redshift $z$ =1.048 \citep{Kedziora-Chudzer-et-al-1997,Kedziora-Chudzer-2001b}. For this line of sight, the transition frequency ($\nu_{\textnormal{o}}$) estimated by the \citet{Cordes-Lazio-2001} model of the Galactic electron distribution is 14.4 GHz. Thus, the Ceduna 6.7 GHz observing frequency is expected to be in the refractive scattering regime ($\nu$ < $\nu_{\textnormal{o}}$) \citep{Turner-et-al-2012}.\par

In this paper, we adopted two techniques to determine the physical characteristics of the source; the ``Earth Orbital Synthesis'' approach which utilizes the fitted parameters from annual cycle fitting \citep{Macquart-Jauncey-2002} and a simple model of an extragalactic point source introduced by \citet{Walker1998}. Modelling using ``Earth Orbital Synthesis'' allows us to estimate the source angular size from the fitted minor axis length scale (i.e. $a_{\textnormal{min}}$ = \textit{L}$\theta_{\textnormal{s}}$, where \textit{L} is the distance of the scattering screen in kpc and $\theta_{\textnormal{s}}$ is the source angular size in microarcsec). We consider this relationship as we assumed that the source size is ``resolved'' by the ISM and it determines the minor axis length scale as changes of both the minor axis length scale and the modulation index appear to be related to the total flux density changes. Therefore, we would expect that the source is most of the time larger than the Fresnel scale regardless of the scattering strength, and larger than refractive scale if we are in strong scattering. \citet{Walker1998} give the scaling of point source modulation index in the regime of refractive scintillation as:

\begin{equation}
m_{\textnormal{p}} = \xi^{- 1/3} = \left(\frac{\nu}{\nu_{\textnormal{o}}}\right)^{17/30}  
\label{eq:mod_index_point_source}
\end{equation}

\noindent
where $m_{\textnormal{p}}$ is the modulation index of a point source (i.e., the rms fractional flux variation), $\xi$ is the scattering strength, $\nu_{\textnormal{o}}$ is the transition frequency and $\nu$ the observing frequency in GHz. Assuming $\nu_{\textnormal{o}}$ = 14.4 GHz as above, the modulation index of a point source and the scattering strength at 6.7 GHz would be expected to be 0.65 and 3.67, respectively. Determination of the refractive angular scale ($\theta_{\textnormal{r}}$) and the size of the first Fresnel zone ($\theta_{\textnormal{F}}$) given in microarcsec can be expressed in the equation~\ref{eq:refractive_Fresnel_scale}:

\begin{equation}
\theta_{\textnormal{r}} = \theta_{\textnormal{ro}} \left(\frac{\nu}{\nu_{\textnormal{o}}}\right)^{-11/5} = \theta_{\textnormal{F}}\xi
\label{eq:refractive_Fresnel_scale}
\end{equation}

\noindent
where the refractive angular scale at the transition frequency ($\theta_{\textnormal{ro}}$) for the Galactic coordinates of interest  = 2.3 $\mu$as \citep{Walker2001}. The refractive angular scale and the size of the first Fresnel zone at $\nu = 6.7$~GHz are then 12.38 and 3.37 $\mu$as, respectively. We then estimated the upper limit of the screen distance as $\theta_{\textnormal{F}}$ = 8/$\sqrt{L\nu}$, where $\theta_{\textnormal{F}}$ is the size of the first Fresnel zone in microarcsec, \textit{L} is the distance of scattering screen in kpc and $\nu$ is the observing frequency in GHz. We estimated the upper limit of the screen distance of PKS\,B1144$-$379 to be 0.84 kpc. \par

\begin{table*}
	\caption{Scintillation parameters and physical properties of PKS\,B1144$-$379 for 2003--2011. Col. 1 gives the distance of scattering screen, Col. 2 the size of first Fresnel zone, Col. 3 the source angular size estimated from the annual cycle fitting with fixed velocity, Col. 4 the scattering strength, Col.5 the brightness temperature, and Col. 6 the Doppler factor.}
	\centering
	\label{table:parameters}
	\begin{tabular}{rrrrrr}
		\hline
		L (kpc) & $\theta_{\textnormal{F}}$ ($\mu$as) & $\theta_{\textnormal{s}}$ ($\mu$as) & $\xi$ & $T_{\textnormal{B}}$ ($10^{14}$ K)  & $\delta$  \\
		\hline
		0.84 &  3.37  & 5.65--15.90    &  1.68 & 0.95--2.74   & 600  \\
		
		     &        &                  &       &                &      \\
		
		0.50 &  4.37  &  9.49--26.73   &  2.17 & 0.34--0.97   & 253  \\
		
		     &        &                  &       &                &      \\
		
		0.10 &  9.78  &  47.45--133.66 &  4.85 & 0.013--0.039 & 17   \\
				
		\hline
	\end{tabular}
\end{table*}

\begin{figure}
	\centering
	\includegraphics[width=\columnwidth]{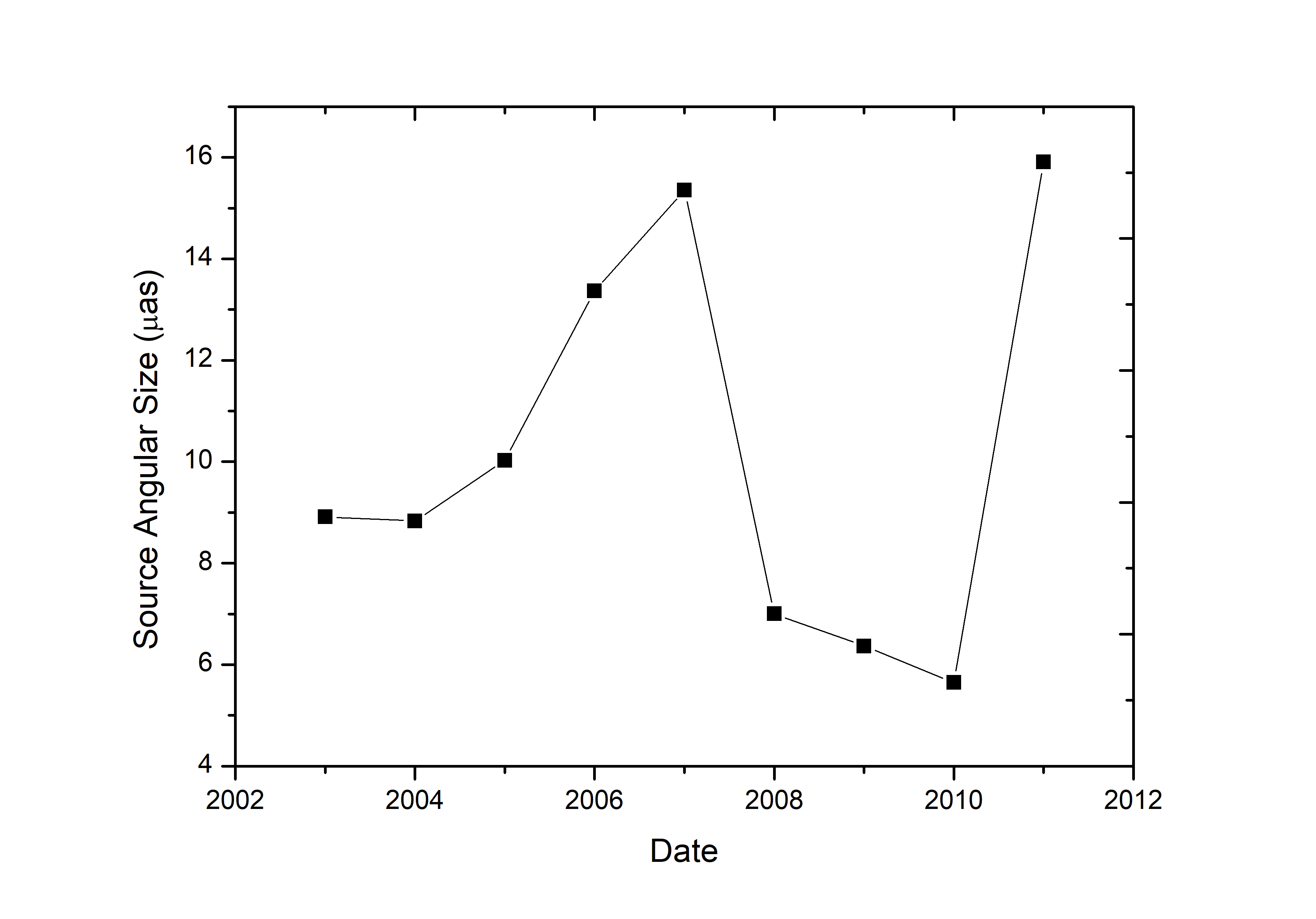}
	\caption{Source angular size estimated using the fitted parameters of the annual cycle model when we fixed the values of both velocities component of scattering screen, $\textbf{\textit{V}}_{\alpha}$ (30 km $s^{-1}$) and $\textbf{\textit{V}}_{\delta}$ (7.7 km $s^{-1}$), between 2003 and 2011. We used 0.84 kpc, the upper limit screen distance to measure the source angular size.}
	\label{fig:angular_size}
\end{figure}

Several possibilities of the screen distance (\textit{L}) and strength of the scattering screen ($\xi$) were considered (refer to Table~\ref{table:parameters}). In our annual cycle fitting, the minor axis scale of the scintillation pattern ($a_{\textnormal{min}}$ = $s_{\textnormal{o}}$ $\sqrt{R}$, where $s_{\textnormal{o}}$ is the characteristic spatial scale of the ISS pattern in km and \textit{R} is the axial ratio) ranged from 0.48--1.95 ($\times 10^6$ km). If instead we use $a_{\textnormal{min}}$ taken from the modelling with the velocities of the screen fixed to be $\textbf{\textit{V}}_{\alpha}$ = 30 km $s^{-1}$ and $\textbf{\textit{V}}_{\delta}$ = 7.7 km $s^{-1}$, the $a_{\textnormal{min}}$ was found to range from 0.71--2.00 ($\times 10^6$ km). For better constraint of fitted parameters, here we only considered the $a_{\textnormal{min}}$ measured from the fixed velocity fits. If the scattering screen is at 0.84 kpc, the Fresnel scale is 3.37 $\mu$as, which is only slightly smaller than the smallest angular size from the annual cycle fitting ($\theta_{\textnormal{s}}$ = 5.65--15.90 $\mu$as). Assuming the source angular size corresponds to the refractive scale and the equation is valid close to the transition frequency, the scattering strength would be 1.68. This would suggest that the scattering is not very strong if the screen is at 0.84 kpc or beyond it. If we move the screen closer, at 500 pc, we have the smallest angular size of 9.49 $\mu$as ($\theta_{\textnormal{s}}$ = 9.49--26.73 $\mu$as), the Fresnel scale would be 4.37 $\mu$as while the scattering strength is 2.17. At screen distance of 100 pc, the smallest angular size is expected to be 47.45 $\mu$as ($\theta_{\textnormal{s}}$ = 47.45--133.66 $\mu$as), the Fresnel scale and the scattering strength would be 9.78 $\mu$as and 4.85, respectively. The screen distance is not well constrained, but a closer screen implies a lower source brightness temperature. We assumed that the screen distance is probably closer than 0.84 kpc or otherwise the scattering strength will be less than 1.68 which seems to be a lot weaker. On the other hand, if the screen distance is closer than 100 pc, this would imply a very strong scattering ($>$ 4.85) in a very nearby screen which is also not expected and the compact fraction of the source would possibly be higher than what we infer from the VLBI data. Variations of timescales longer than a day for PKS\,B1144$-$379 also indicate that more distant screen ($>$ 100 pc) is expected as measured by \citet{Turner-et-al-2012}. In addition, a handful of rapid IDV sources scintillating through nearby screens have been argued to be scintillating in weak scattering at frequencies as high as 6.7~GHz \citep[e.g.][]{Kedziora-Chudzer-et-al-1997}. For PKS\,B1144$-$379, the long timescale of variation and the annual cycle fit constrains the characteristic length scale of the scintillation pattern to $>$ $10^{6}$ km. For the more rapid scintillators that have had an annual cycle measured (PKS\,1257$-$326 and J1819$+$3845), the scintillation pattern scale is more than the order of magnitude smaller, and the scattering screen is inferred to be within tens of parsec. If PKS\,B1144$-$379 is scintillating through such a nearby screen, the scintillation pattern scale must be much larger than the Fresnel scale, implying either the source angular size is much larger or the scattering is very strong (leading to a larger refractive length scale). Such strong scattering on the nearby screen is not expected (or at least would have been unprecedented). Figure~\ref{fig:angular_size} shows the source angular size estimated using the fitted parameters of the annual cycle model when we constrained the values of both velocities component of scattering screen, $\textbf{\textit{V}}_{\alpha}$ (30 km $s^{-1}$) and $\textbf{\textit{V}}_{\delta}$ (7.7 km $s^{-1}$) for the years 2003--2011. Here we used 0.84 kpc, the upper limit screen distance to measure the source angular size. The source angular size of PKS\,B1144$-$379 determined using the ``Earth Orbital Synthesis'' ranged from 5.65--15.90 $\mu$as.   \par

\begin{figure}
	\centering
	\includegraphics[width=\columnwidth]{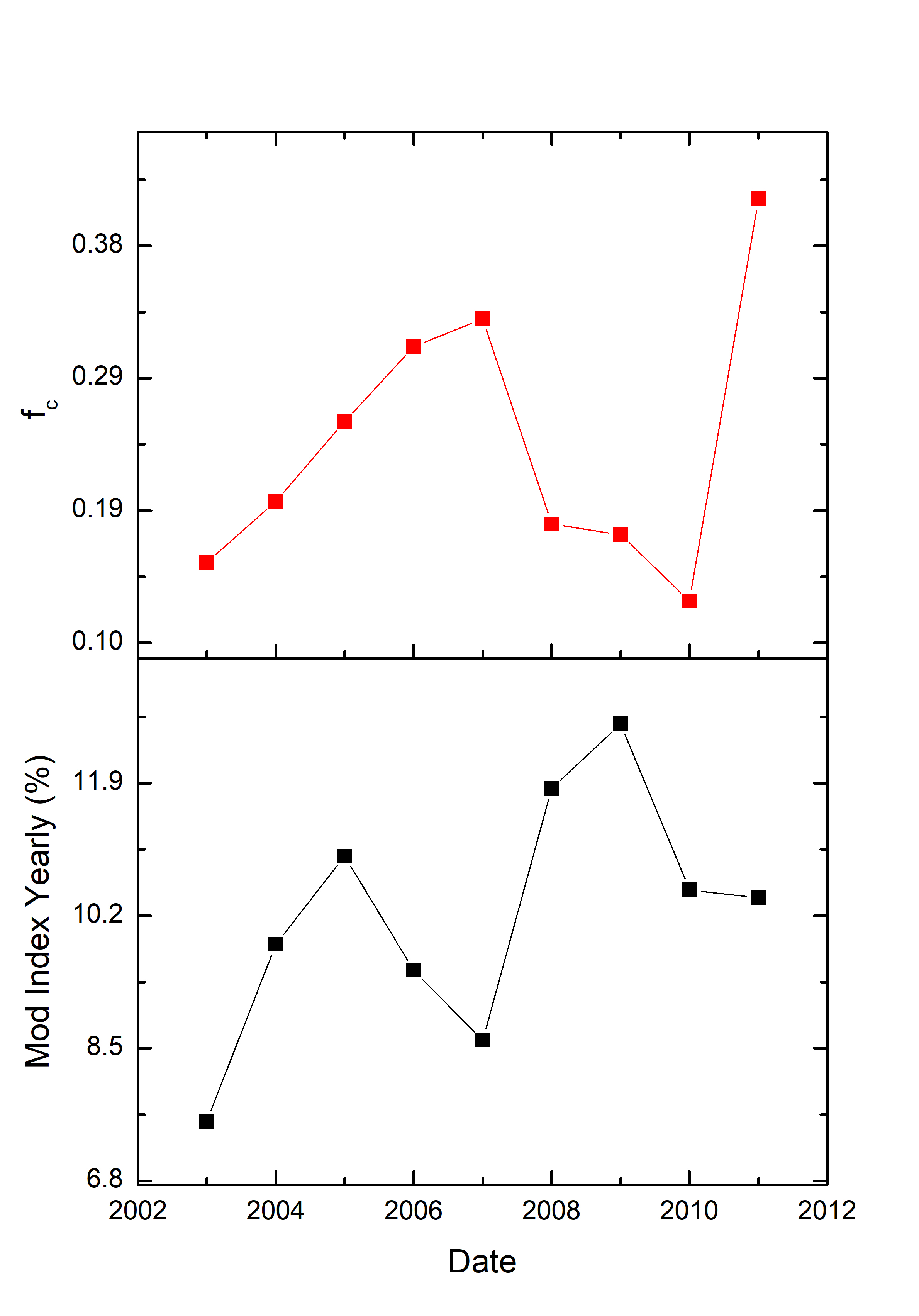}
	\caption{Comparison between compact fraction $f_{\textnormal{c}}$ (top panel) and averaged modulation index of PKS\,B1144$-$379 (bottom panel) between 2003 and 2011.}
	\label{fig:compact_fraction}
\end{figure}

The simple model of an extragalactic point source introduced by \citet{Walker1998} assumes that all the flux density is in the scintillating component. However, observations typically show that only a fraction of the total flux density is time variable with some of the emission arising in a more extended milliarcsecond-scale jet. Furthermore, the source angular size determined from this approach is derived using only the modulation index and the assumption that the transition frequency is 14.4 GHz. This approach does not utilise information obtained from the modelling of the annual cycle in the modulation timescale (known as ``Earth Orbital Synthesis'' ), which also yields information on the source angular size. The modulation index has been used by various authors to characterise the linear scale of the region responsible for the intraday variability \citep{Quirrenbach-et-al-1992,Peng-et-al-2000,Kedziora-Chudzer-2001b}. Here, we altered the equation of refractive scintillation given by \citet{Walker1998} and to include the compact fraction of the source ($f_{\textnormal{c}}$ = $S_{\textnormal{core}}$/$S_{\textnormal{total}}$) as:

\begin{equation}
m = m_{\textnormal{p}} \left(\frac{\theta_{\textnormal{r}}}{\theta_{\textnormal{s}}}\right)^{7/6}  f_{\textnormal{c}}
  = \xi^{-1/3} \left(\frac{\xi \times \theta_{\textnormal{F}}}{\theta_{\textnormal{s}}}\right)^{7/6} f_{\textnormal{c}}
\label{eq:source_size}
\end{equation}

\noindent
where $\textit{m}$ is the observed modulation index, $m_{\textnormal{p}}$ is the modulation index of point source (0.65), $\theta_{\textnormal{r}}$ the refractive scale (12.38 $\mu$as), $\theta_{\textnormal{F}}$ the size of first Fresnel zone (3.37 $\mu$as), $\theta_{\textnormal{S}}$ the source angular size estimated from the annual cycle fitting with fixed velocity, $\xi$ the scattering strength of screen distance of 0.84 kpc and $f_{\textnormal{c}}$ the compact fraction. Figure~\ref{fig:compact_fraction} compares the compact fraction measured using the Equation~\ref{eq:source_size} and the averaged modulation index of PKS\,B1144$-$379. The source compact fraction was found to be increased where the total flux density peaks and the averaged modulation index is reduced (2006, 2007 and 2011). This might indicates that the scintillating component has brightened and expanded. \par

\begin{figure}
	\centering
	\includegraphics[width=\columnwidth]{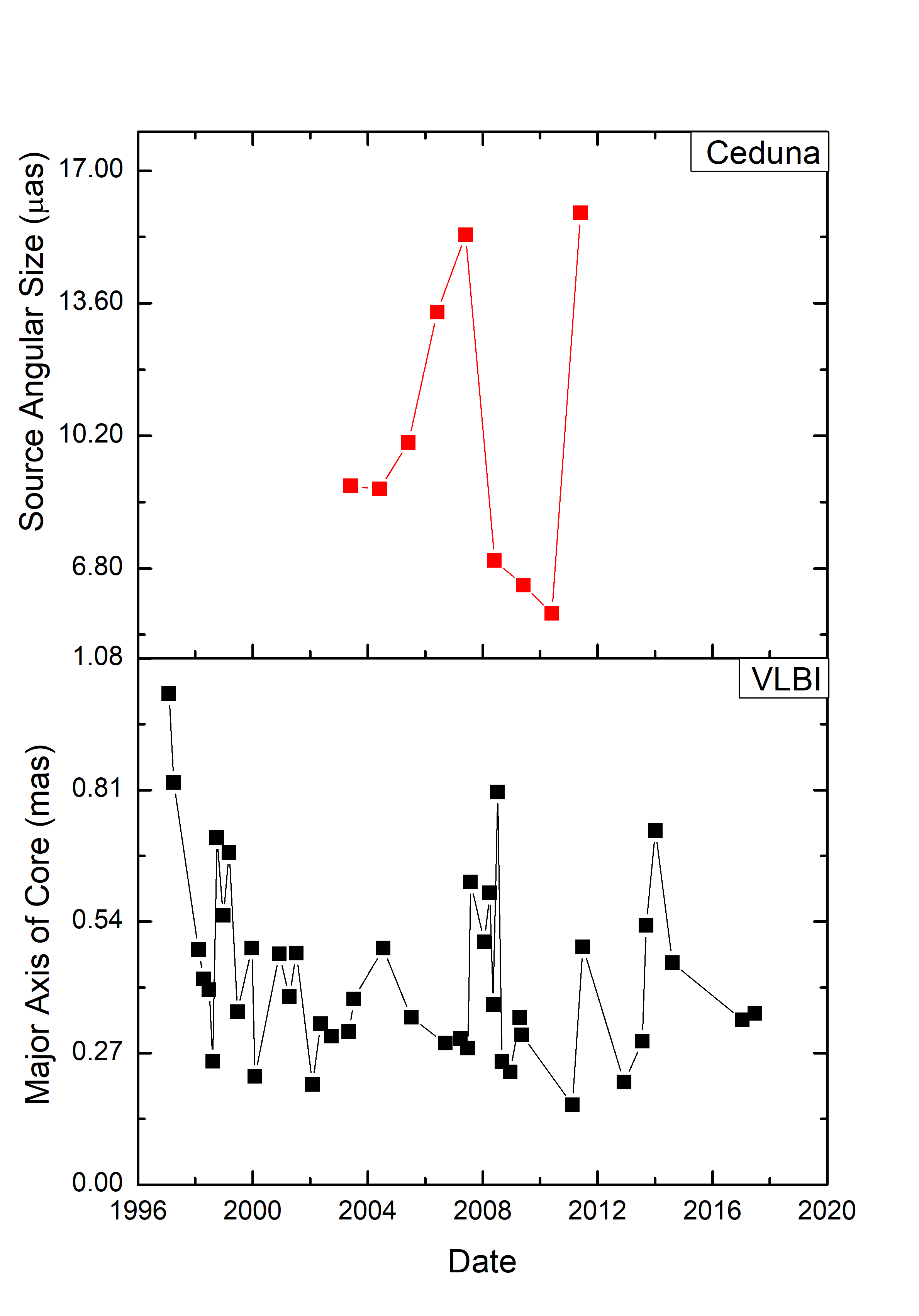}
	\caption{Comparison between source angular size estimated using the  ``Earth Orbital Synthesis'' of 6.7 GHz COSMIC data (top panel) and major axis of core component measured from the 8.6 GHz VLBI radio images (bottom panel). In the period 2001 to 2013, the major axis of the core component was estimated to vary between 0.16 and 0.81 mas. The VLBI radio images taken from The Radio Reference Frame Image Database (RRFID) and The Astrogeo VLBI database.}
	\label{fig:angular_size_VLBI_Ceduna}
\end{figure}

\begin{figure*}
	\centering
	\includegraphics[width=\textwidth]{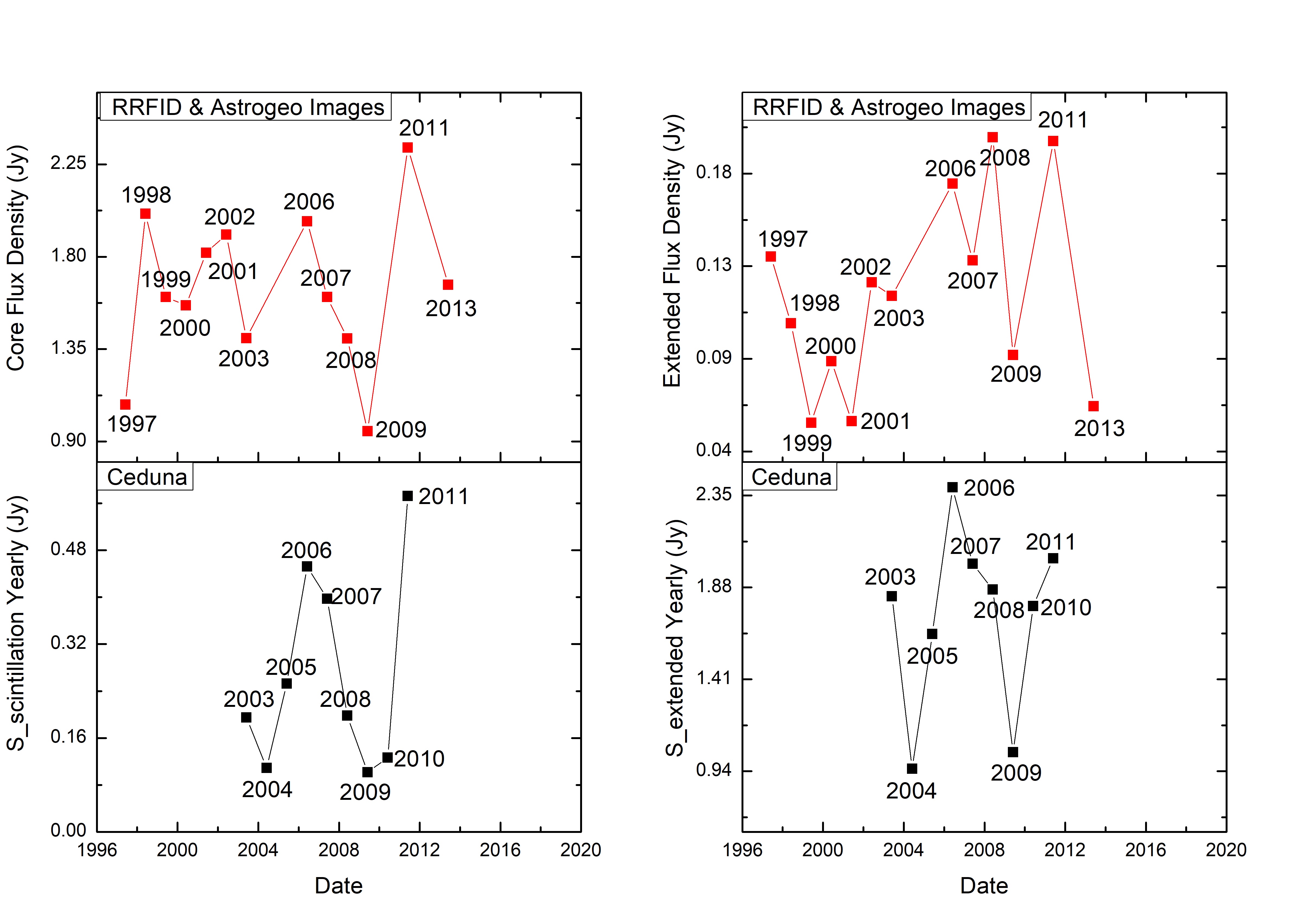}
	\caption{Comparison of the flux density of scintillating component (left panel) and extended area (right panel) measured from COSMIC data (bottom panel) and the 8.6 GHz VLBI radio images (top panel). The flux density of the scintillating component and non-scintillating components from the COSMIC data were measured by using the angular size estimated from the annual cycle. The flux density of the scintillating and non-scintillating components were found to range between 0.10--0.57 Jy and 0.95--2.39 Jy, respectively. Core flux density were measured to range between 0.95--2.33 Jy whereas the extended component varied from 0.06--0.19 Jy.}
	\label{fig:flux_scintillation_extended}
\end{figure*}

In addition to the COSMIC dataset presented in this paper, information on the physical characteristics of the source can also be determined from VLBI radio images. Figure~\ref{fig:angular_size_VLBI_Ceduna} compares the source angular size estimated using the ``Earth Orbital Synthesis'' of the COSMIC data at 6.7 GHz and major axis of core component measured from the 8.6 GHz VLBI radio images. The VLBI radio images were obtained from The Radio Reference Frame Image Database (RRFID) (https://www.usno.navy.mil/USNO/astrometry/vlbi-products/rrfid) and the Astrogeo VLBI FITS image database (http://astrogeo.org/vlbi-images). The upper limit of the VLBI core size was measured from an elliptical Gaussian model fit to the image data made using the DIFMAP software package \citep{Shepherd1997}. In the period 2001 to 2013, the major axis of the core component was estimated to vary between 0.16 and 0.81 mas. We found quite similar pattern between these two datasets with the source angular size estimated from the COSMIC data modulation index lagging the value obtained from the major axis of the core component of VLBI radio images. The core appears to expand first at the scales of ISS, and then later at VLBI scales implying a new component moving out in the jet. Figure~\ref{fig:flux_scintillation_extended} compares the flux density of both the compact and extended emission regions for PKS\,B1144$-$379 estimated through two independent techniques. From the COSMIC data we can measure the flux density of both the scintillating and non-scintillating (extended) components, while from the VLBI images we have the flux density of the core component compared to the total flux density of the source.  

\begin{figure}
	\centering
	\includegraphics[width=\columnwidth]{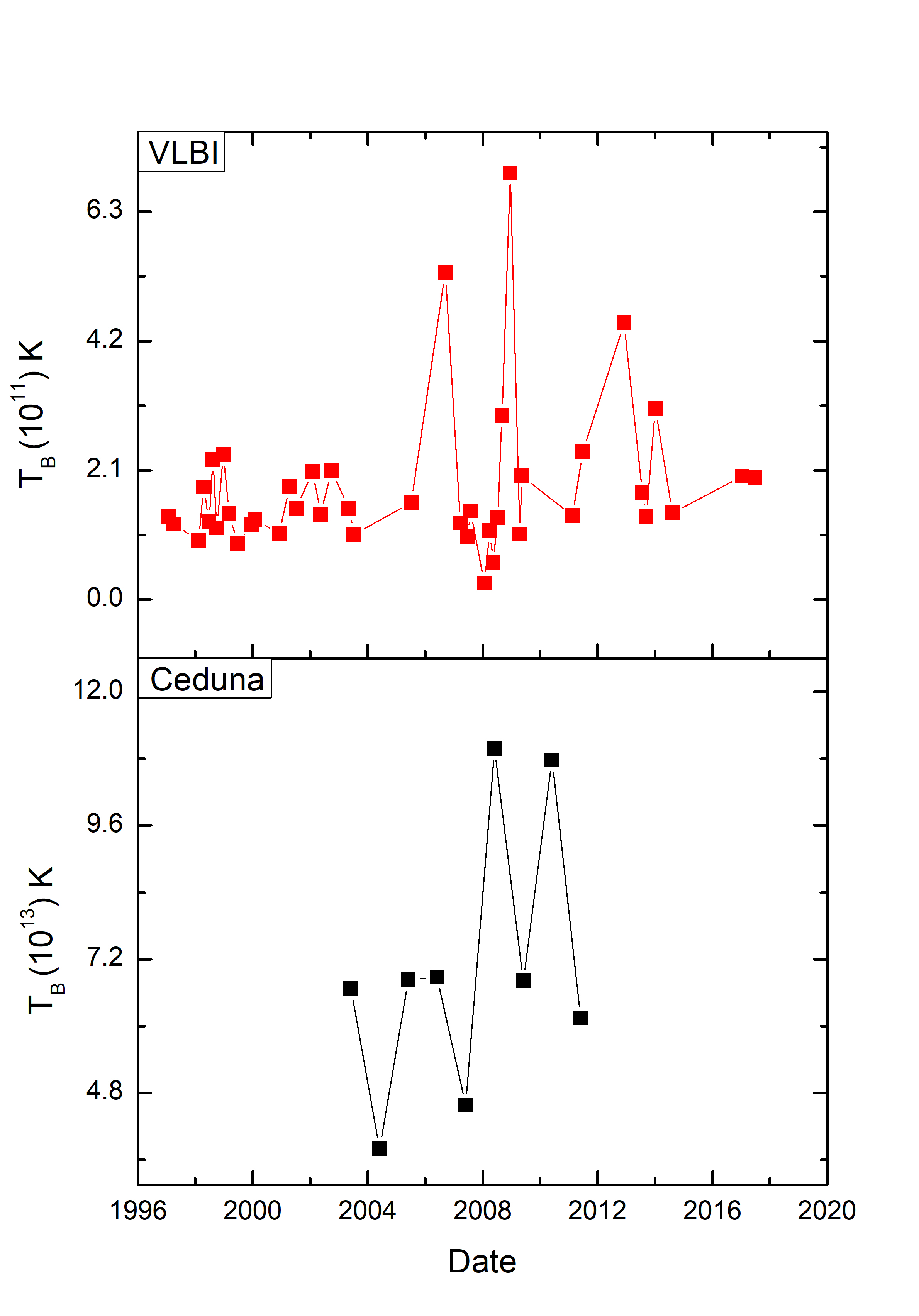}
	\caption{Comparison between brightness temperature estimated from the 8.6 GHz VLBI radio images (top panel) and using the source angular size of 6.7 GHz COSMIC data (bottom panel). The source brightness temperature measured from the COSMIC data ranged from $3.8\times10^{13}$ to $1.1\times10^{14}$ K, whereas the lower limit of the source brightness temperature estimated from the VLBI radio images were found to vary between $2.7\times10^{10}$ and $6.9\times10^{11}$ K. The VLBI radio images taken from The Radio Reference Frame Image Database (RRFID) and The Astrogeo VLBI database.}
	\label{fig:brightness_temperature}
\end{figure}

The modulation index of the scintillating component ($m_{\textnormal{scint}}$) is expected to be: 

\begin{equation}
m_{\textnormal{scint}} = RMS/S_{\textnormal{scint}}  = m_{\textnormal{p}}\left(\frac{\theta_{\textnormal{r}}}{\theta_{\textnormal{s}}}\right)^{7/6} 
\label{eq:scintillation_mod_index}
\end{equation} 

\noindent
where the RMS values are the yearly-averaged quantities, $\theta_{\textnormal{r}}$ the refractive angular size (12.38 $\mu$as), $\theta_{\textnormal{s}}$ from the annual cycle fitting (minor axis of the length scale), and $\textit{m}_{\textnormal{p}}$ based on an assumed transition frequency of 14.4 GHz. The flux density of the scintillating and non-scintillating components were found to vary by a factor of 8 and 3, respectively. The estimated flux density in the scintillating and non-scintillating components correlates with the flux densities measured from the VLBI radio images. Figure~\ref{fig:brightness_temperature} shows the source brightness temperature estimated from the source angular size of COSMIC data ($S = 2.65T_{\textnormal{B}}\theta^{2}$/$\lambda^{2}$, where $S$ is the flux density of scintillating component in Jy, $\theta$ the source angular size measured from the annual cycle fitting in arcmin, $\lambda$ the observing wavelength in cm and $T_{\textnormal{B}}$ brightness temperature in K) and a lower limit for the source brightness temperature measured from the VLBI radio images \citep[method adopted from][]{Frey-et-al-2015}. The source brightness temperature measured from the COSMIC data ranged from $3.8\times10^{13}$ to $1.1\times10^{14}$ K (assuming that the observed variations are due to scintillation), whereas the lower limit of the source brightness temperature estimated from the VLBI radio images were found to vary between $2.7\times10^{10}$ and $6.9\times10^{11}$ K.  The source brightness temperature inferred from the COSMIC data exceeds the inverse Compton limit \citep[$\sim 10^{12}$ K;][]{Kellermann-Pauliny-1969} and implies a Doppler boosting factor of $>$ 100. Estimation of Doppler factor was under the assumption of brightness temperature limited incoherent synchrotron emission \citep{Marscher1998}. The inference of brightness temperatures that violate the inverse Compton limit strengthens the case that the radio IDV observed in PKS\,B1144$-$379 is caused by interstellar scintillation. The estimated brightness temperature in this paper is in good agreement with the brightness temperature measured by \citet{Turner-et-al-2012}. They found that the source brightness temperature for PKS\,B1144$-$379 to be $6.2\times10^{12}$ K at 4.9 GHz, under the assumption that the observed flux density variations were due to scintillation. \citet{Rickett-et-al-2006} reported that the majority of scintillating AGN have maximum brightness temperatures of $10^{11}-10^{12}$ K.   \par

As previously mentioned, a closer screen implies a lower brightness temperature. We deduced the brightness temperature and inferred Doppler factor are extremely high if the screen is at 0.84 kpc. Table~\ref{table:parameters} shows several possibilities of the screen distance (L) that we considered to estimate the brightness temperature and the Doppler factor. Equation~\ref{eq:scintillation_mod_index} used to measure the flux density of scintillating component and we assumed the source angular size corresponds to the refractive scale. If the screen is at 0.84 kpc, the brightness temperature would range from $9.48 \times 10^{13}$  to $2.74 \times 10^{14}$ K  and this would suggest Doppler factor of $\sim$ 600. If we move the screen closer, at 500 pc, the brightness temperature could be lowered down to 3.36--9.72 ($10^{13}$ K) while the Doppler factor is expected to be 253. At a screen distance of 100 pc, the brightness temperature and the Doppler factor would be 1.34--3.89 ($10^{12}$ K) and 17, respectively. A reasonable value of brightness temperature and inferred Doppler factor imply that the screen distance of PKS\,B1144$-$379 is probably closer than 0.84 kpc. \par

\section{Discussion}

The COSMIC data shows consistent annual modulations in the timescale of variability of PKS\,B1144$-$379 from 2003 through 2011. This represents strong evidence that the radio IDV in PKS\,B1144$-$379 is due to interstellar scintillation rather than any intrinsic mechanism. The longer characteristic timescales for PKS\,B1144$-$379 compared to most other long-term monitoring radio IDV sources studied to date, implies that the scattering screen along the line of sight is further away (upper limit of $\approx$ 0.84 kpc) compared to more rapidly scintillating sources. A nearby scattering screen in which the scattering phenomenon occurs in a very local region of turbulence is inferred for a number of the rapid scintillating sources. For example, \citet{Bignall-et-al-2003} estimate the scattering screen for PKS\,B1257$-$326 to be at a distance of 12 $\pm$ 4 pc at 4.8 GHz and 15 $\pm$ 6 pc at 8.6 GHz. Similarly, \citet{Dennett-Bruyn-2003} estimate the scattering screen for PMN\,J1819$-$3845 to be between 1--12 pc away, while \citet{Rickett-et-al-2002} find the scattering screen distance to be 25 pc for PKS\,0405$-$385. For most extragalactic sources, the relationship between the characteristic timescale and the screen distance lies between a square root and a linear dependence \citep{Rickett-et-al-2002}. The angular size required for a source to scintillate in the ISM is governed by the angular size of the first Fresnel zone. Those sources which show larger variations with longer characteristic timescales are more likely located behind scattering screens at much larger distances \citep{Rickett-et-al-1995}. The smaller angular sizes inferred for longer characteristic timescale scintillators thus correspond to higher brightness temperatures.  \par

The characteristic timescale of the flux density variations in PKS\,B1144$-$379 appear to be affected by source evolution. Long-term monitoring of variations of the flux density show that the source has experienced a parsec-scale outburst which is commonly seen in many flat spectrum AGN \citep{Lister-et-al-2009}. The loose correlation observed between the characteristic timescale and the major axis of the core component measured from the 8.6 GHz VLBI radio images suggests that there is a relationship between the source evolution and properties of the radio IDV. \citet{Kellermann-Pauliny-1981} reported that flat-spectrum radio sources often show outbursts on timescales of months to years at centimetre wavelengths and these intrinsic variations are associated with the relativistic jets observed toward these sources in VLBI images. For PKS\,B1257$-$326, a parsec-scale outburst was detected that produced no significant change in the scintillating flux density. The hypothesis is that the slowly varying source component has an optically thick spectrum and the $\mu$as core does not lie behind what is presumably an expanding parsec-scale jet, as might be expected if the scintillating component were the core of the base of the parsec-scale jet \citep{Bignall-et-al-2003}. For PKS\,B1144$-$379 there is a significant change in the scintillating flux density of the source, hence the $\mu$as core might lie behind the expanding parsec-scale jet and the outburst is probably occurring in the same emitting region as the scintillating core. Possible consequences of flares on radio IDV have been studied by \citet{Liu-et-al-2013} for the source 1156+295 at 4.8 GHz. Changes in the radio IDV characteristic timescale was compared with the evolution of the 43 GHz VLBA core size. \citet{Liu-et-al-2013} hypothesized that the very low variability amplitudes measured for some periods for this source may be the consequence of an increased source size related to the flare.  \par

We have used the ``Earth Orbital Synthesis'' technique to study both the source structure and the ISM of the scattering screen. This method also can be used to investigate the polarization structure \citep{Jauncey-Macquart-2001}. However, there are some limitations of ``Earth Orbital Synthesis'' that need to be considered. The accuracy of the method depends on the measurement of the characteristic timescale and also assumes a single, stable scattering screen. Sparse measurements of the characteristic timescale increase the uncertainty in the determination of the source and screen parameters returned by annual cycle fitting. The assumption of an unchanging scattering screen over the course of the observations is also not necessarily true. For example, the scattering medium that was responsible for the strong and rapid radio IDV of J1819$+$3845 has moved away, leading to a significant decrease of the variability timescales \citep{Macquart-Bruyn-2007,Koay-et-al-2011}. Measurement of source sizes using this approach are directly dependent on the distance to the scattering screen, as a closer scattering screen implies a larger source angular size. In our study, changes in the scintillation length scale in successive annual cycles appear to be related to the long term intrinsic flux density changes in the source. We also made an assumption that there is no change in average scattering properties over the course of the observations. \par

Smaller values of reduced $\chi^{2}$ suggest that the anisotropic ISS model is more accurate in its fitting of the annual cycle in the characteristic timescale of PKS\,B1144$-$379. Nevertheless, for some of the observing years, the fitted parameters are not well constrained, with the main cause thought to be due to the influence of source evolution on the characteristic timescale. Large anisotropy ratios were inferred from the fits for some years (2005 and 2010), with the axial ratio of the scattering screen going to zero (infinity) and both components of the relative transverse velocity of the scintillation pattern and the observer tending towards unrealistic values. The fitted parameters were found to be better behaved when we fixed the values of both velocity components of the scattering screen ($\textbf{\textit{V}}_{\alpha}$ and $\textbf{\textit{V}}_{\delta}$ were fixed at 30 and 7.7 km $s^{-1}$, respectively).   \par

Some previous studies of radio IDV have used anisotropic ISS models to fit the characteristic timescale.  The presence of a magnetic field in the plasma can lead to an anisotropic scattering medium \citep{McCallum2009}.  The effect of the anisotropic medium on the scintillation is greatest when the scattering screen is thin and the magnetic field must be well ordered through the scattering screen (i.e along the line of sight) \citep{Dennett-Bruyn-2003,Rickett-Coles-2004}. In their study of PKS\,B1257$-$326, \citet{Bignall-et-al-2003,Bignall-et-al-2006} found evidence for anisotropic scattering and a highly elongated scintillation pattern for both the 4.9 and 8.5 GHz observations, with axial ratios ($\textit{R}$) greater than 10:1 extended in a north-west direction on the sky. The 2-D anisotropic model led them to conclude that the high anisotropy of the scintillation pattern caused degenerate solutions for the scintillation velocity. \citet{Walker-et-al-2009} made a comparison between the 1-D and 2-D model of anisotropy for PKS\,B1257$-$326 and J1819+3845. They found that for J1819+3845, the optimum fit is infinite anisotropy, but not for PKS\,1257$-$326 where a 2-D model with an axis ratio of 6 fits the data significantly better than the 1-D model. This suggests that for J1819+3845, it requires extremely anisotropic scattering and that infers a region where the magnetic field is both strong and highly ordered. They also claimed that in the 1-D scintillation model, the anisotropy can be as high as $10^{5}$. Other radio IDV sources like PKS\,1322$-$110 \citep{Bignall-et-al-2019} and PKS\,B0405$-$385 \citep{Rickett-et-al-2002} have also been argued to show scintillation consistent with anisotropic scattering.  \par

Anisotropic scattering could be due to either the scattering screen or anisotropy of the emitting region \citep{Gabanyi-et-al-2009}. In order to determine the cause of anisotropy in PKS\,B1144$-$379, we compared the position angle of the anisotropy (measured by $90^{o}$ - $\beta$) with the position angle of the VLBI-scale jet. Over a 20 year period (1997 January 30 and 2017 June 28), the position angle of the VLBI jet in PKS\,B1144$-$379 has been determined using images from the Radio Reference Frame Image Database (RRFID) and the Astrogeo VLBI FITS Image databases. The jet position angles varied between $171^{o}$ to $-145^{o}$. The VLBI-scale jet measurements show good agreement with the TANAMI program observed by \citet{Ojha-et-al-2010}. The position angle of the anisotropy estimated from the annual cycle fitting of Ceduna data observed from 2003 through 2011 ranged from 18 to 56 degrees. The difference in the position angle between the scintillation and the VLBI-scale jet suggests that the anisotropic scattering is due to the ISM rather than the source structure. The ``overshoot'' of the structure function which exceeds the saturation level (twice variance) is also consistent with anisotropy in the scattering screen. As reported by \citet{Rickett-et-al-2002}, the negative $"$overshoot$"$ in the AutoCorrelation Function (ACF) data of PKS\,0405$-$385 during its episodic periods of fast scintillation indicates that the scattering medium is highly anisotropic. They argued that the negative ``overshoot'' is an effect of anisotropic scattering in the ISM, and unlikely to be produced by anisotropy of source structure.  \par

The origin of the scattering screens of extreme scintillators is still a mystery. Previous models proposed that the scattering screen might be produced by the current sheets seen edge-on \citep{Goldreich-Sridhar-2006,Pen-Levin-2014} or the ionized skins of tiny molecular clouds \citep{Walker1998}, but the real cause of this phenomenon is still questionable. \citet{Walker-et-al-2017} have proposed that the scattering structures that cause radio IDV are associated with hot stars. They found under the condition of a highly-anisotropic (i.e. long and thin) scattering screen, the long axis of the screen derived from the data points at a nearby star, the screen velocity perpendicular to the long axis roughly matches the velocity of the star in the same direction, and the screen distance also matches that of the star. \citet{Walker-et-al-2017} found a probability of this association occurring by chance in their two cases (J1819$+$3845 and PKS\,1257$-$326) of $10^{-7}$. Several authors have reported the association of hot star with the origin of scattering screen. For example, B star Spica is associated with the scattering screen of PKS\,1322$-$110 \citep{Bignall-et-al-2019}, \citet{Dennett-deBruyn-2002} suggested the nearby A star Vega might be associated with J1819$+$3845, and Alhakim ($\iota$ Centauri) for PKS\,1257$-$326 \citep{Walker1998}.  In this study, we used the Gaia catalogue to search for nearby hot stars that can be associated with the scattering screen for PKS\,B1144$-$379. We found two A stars (Gaia DR2 5384837413688122624 and Gaia DR2 3463396279568792576) which have transverse velocity ($\sim$ 22 km$s^{-1}$), direction and distance ($\sim 7.4 \times 10^{2}$ pc) roughly matching the velocity of the scattering screen perpendicular to the line of sight ($\textbf{\textit{V}}_{\textnormal{perp}}$) measured from the fitted parameters of the unweighted anisotropic annual cycle with fixed values of $\textbf{\textit{V}}_{\alpha}$ (30 km$s^{-1}$) and $\textbf{\textit{V}}_{\delta}$ (7.7 km$s^{-1}$). However, further investigation is needed to verify this finding. \par

\section{Conclusions}

We have analysed the variations of flux density of PKS\,B1144$-$379 using the 6.7 GHz COSMIC dataset, the Astrogeo VLBI FITS Image database and the United States Naval Observatory (USNO) Radio Reference Frame Image Database (RRFID) project. We find that the intraday variability of PKS\,B1144$-$379 is well explained as originating from interstellar scintillation (ISS). We have independently fitted an annual cycle model for each of the years 2003 to 2011 inclusive and find that it suggests the scattering screen has an anisotropic structure and that evolution of the source structure influences the interstellar scintillation pattern. From the yearly changes in characteristic timescale of scintillation, we infer changes in the size of the microarcsecond-scale, scintillating component of the source. Changes in the source inferred from the scintillation measurements are also consistent with VLBI observations, although scintillation probes a much smaller angular scale. The component was found to be at its most compact during two flares in total flux density (2005 and 2008).  \par

\section*{Acknowledgements}

We thank Bill Coles and members of the ATESE project for their helpful suggestions in improving this paper. This research has made use of material from The Astrogeo VLBI FITS Image database and The United States Naval Observatory (USNO) Radio Reference Frame Image Database (RRFID) projects.

\section*{Data availability}

The data underlying this article are available in the article and in its online supplementary material.

\bsp	
\label{lastpage}

\bibliographystyle{mnras}
\bibliography{ref}

\begin{thebibliography}{}
\makeatletter
\relax
\def\mn@urlcharsother{\let\do\@makeother \do\$\do\&\do\#\do\^\do\_\do\%\do\~}
\def\mn@doi{\begingroup\mn@urlcharsother \@ifnextchar [ {\mn@doi@}
  {\mn@doi@[]}}
\def\mn@doi@[#1]#2{\def\@tempa{#1}\ifx\@tempa\@empty \href
  {http://dx.doi.org/#2} {doi:#2}\else \href {http://dx.doi.org/#2} {#1}\fi
  \endgroup}
\def\mn@eprint#1#2{\mn@eprint@#1:#2::\@nil}
\def\mn@eprint@arXiv#1{\href {http://arxiv.org/abs/#1} {{\tt arXiv:#1}}}
\def\mn@eprint@dblp#1{\href {http://dblp.uni-trier.de/rec/bibtex/#1.xml}
  {dblp:#1}}
\def\mn@eprint@#1:#2:#3:#4\@nil{\def\@tempa {#1}\def\@tempb {#2}\def\@tempc
  {#3}\ifx \@tempc \@empty \let \@tempc \@tempb \let \@tempb \@tempa \fi \ifx
  \@tempb \@empty \def\@tempb {arXiv}\fi \@ifundefined
  {mn@eprint@\@tempb}{\@tempb:\@tempc}{\expandafter \expandafter \csname
  mn@eprint@\@tempb\endcsname \expandafter{\@tempc}}}

\bibitem[\protect\citeauthoryear{{Abdo} et~al.,}{{Abdo}
  et~al.}{2009}]{Abdo-et-al-2009}
{Abdo} A.~A.,  et~al., 2009, \mn@doi [\apj] {10.1088/0004-637X/700/1/597},
  \href {http://adsabs.harvard.edu/abs/2009ApJ...700..597A} {700, 597}

\bibitem[\protect\citeauthoryear{Azami, Mohammadi  \& Bozorgtabar}{Azami
  et~al.}{2012}]{Azami-et-al-2012}
Azami H.,  Mohammadi K.,   Bozorgtabar B.,  2012, Journal of Signal and
  Information Processing, 3, 39

\bibitem[\protect\citeauthoryear{{Baars}, {Genzel}, {Pauliny-Toth}  \&
  {Witzel}}{{Baars} et~al.}{1977}]{Baars-et-al-1977}
{Baars} J.~W.~M.,  {Genzel} R.,  {Pauliny-Toth} I.~I.~K.,   {Witzel} A.,  1977,
  \aap, \href {http://adsabs.harvard.edu/abs/1977A%26A....61...99B} {61, 99}

\bibitem[\protect\citeauthoryear{{Bignall} et~al.,}{{Bignall}
  et~al.}{2003}]{Bignall-et-al-2003}
{Bignall} H.~E.,  et~al., 2003, \mn@doi [\apj] {10.1086/346180}, \href
  {https://ui.adsabs.harvard.edu/abs/2003ApJ...585..653B} {585, 653}

\bibitem[\protect\citeauthoryear{{Bignall}, {Macquart}, {Jauncey}, {Lovell},
  {Tzioumis}  \& {Kedziora-Chudczer}}{{Bignall}
  et~al.}{2006}]{Bignall-et-al-2006}
{Bignall} H.~E.,  {Macquart} J.~P.,  {Jauncey} D.~L.,  {Lovell} J.~E.~J.,
  {Tzioumis} A.~K.,   {Kedziora-Chudczer} L.,  2006, \apj, 652, 1050

\bibitem[\protect\citeauthoryear{{Bignall} et~al.,}{{Bignall}
  et~al.}{2019}]{Bignall-et-al-2019}
{Bignall} H.,  et~al., 2019, \mn@doi [\mnras] {10.1093/mnras/stz1559}, \href
  {https://ui.adsabs.harvard.edu/abs/2019MNRAS.487.4372B} {487, 4372}

\bibitem[\protect\citeauthoryear{{Blanchard}}{{Blanchard}}{2013}]{Blanchard2013}
{Blanchard} J.,  2013, PhD thesis, University of Tasmania

\bibitem[\protect\citeauthoryear{{Bolton} \& {Shimmins}}{{Bolton} \&
  {Shimmins}}{1973}]{Bolton-Shimmins-1973}
{Bolton} J.~G.,  {Shimmins} A.~J.,  1973, Australian Journal of Physics
  Astrophysical Supplement, \href
  {http://adsabs.harvard.edu/abs/1973AuJPA..30....1B} {30, 1}

\bibitem[\protect\citeauthoryear{{Carter}}{{Carter}}{2008}]{Carter2008}
{Carter} S.,  2008, PhD thesis, University of Tasmania

\bibitem[\protect\citeauthoryear{{Carter}, {Ellingsen}, {Macquart}  \&
  {Lovell}}{{Carter} et~al.}{2009}]{Carter-at-al-2009}
{Carter} S.~J.~B.,  {Ellingsen} S.~P.,  {Macquart} J.~P.,   {Lovell} J.~E.~J.,
  2009, \mn@doi [\mnras] {10.1111/j.1365-2966.2009.14824.x}, 396, 1222

\bibitem[\protect\citeauthoryear{{Cordes} \& {Lazio}}{{Cordes} \&
  {Lazio}}{2001}]{Cordes-Lazio-2001}
{Cordes} J.~M.,  {Lazio} T.~J.~W.,  2001, \mn@doi [\apj] {10.1086/319442},
  \href {http://adsabs.harvard.edu/abs/2001ApJ...549..997C} {549, 997}

\bibitem[\protect\citeauthoryear{{Dennett-Thorpe} \& {de
  Bruyn}}{{Dennett-Thorpe} \& {de Bruyn}}{2000}]{Dennett-Thorpe-de-Bruyn-2000}
{Dennett-Thorpe} J.,  {de Bruyn} A.~G.,  2000, \mn@doi [\apjl]
  {10.1086/312459}, \href {http://adsabs.harvard.edu/abs/2000ApJ...529L..65D}
  {529, L65}

\bibitem[\protect\citeauthoryear{{Dennett-Thorpe} \& {de
  Bruyn}}{{Dennett-Thorpe} \& {de Bruyn}}{2001}]{Dennett-Bruyn-2001}
{Dennett-Thorpe} J.,  {de Bruyn} A.~G.,  2001, in {Laing} R.~A.,  {Blundell}
  K.~M.,  eds,  Astronomical Society of the Pacific Conference Series Vol. 250,
  Particles and Fields in Radio Galaxies Conference. p.~133

\bibitem[\protect\citeauthoryear{{Dennett-Thorpe} \& {de
  Bruyn}}{{Dennett-Thorpe} \& {de Bruyn}}{2002}]{Dennett-deBruyn-2002}
{Dennett-Thorpe} J.,  {de Bruyn} A.~G.,  2002, \mn@doi [\nat]
  {10.1038/415057a}, \href {http://cdsads.u-strasbg.fr/abs/2002Natur.415...57D}
  {415, 57}

\bibitem[\protect\citeauthoryear{{Dennett-Thorpe} \& {de
  Bruyn}}{{Dennett-Thorpe} \& {de Bruyn}}{2003}]{Dennett-Bruyn-2003}
{Dennett-Thorpe} J.,  {de Bruyn} A.~G.,  2003, \mn@doi [\aap]
  {10.1051/0004-6361:20030329}, \href
  {https://ui.adsabs.harvard.edu/abs/2003A&A...404..113D} {404, 113}

\bibitem[\protect\citeauthoryear{{Frey}, {Paragi}, {Fogasy}  \&
  {Gurvits}}{{Frey} et~al.}{2015}]{Frey-et-al-2015}
{Frey} S.,  {Paragi} Z.,  {Fogasy} J.~O.,   {Gurvits} L.~I.,  2015, \mn@doi
  [\mnras] {10.1093/mnras/stu2294}, \href
  {https://ui.adsabs.harvard.edu/abs/2015MNRAS.446.2921F} {446, 2921}

\bibitem[\protect\citeauthoryear{{Fuhrmann} et~al.,}{{Fuhrmann}
  et~al.}{2002}]{Fuhrmann-et-al-2002}
{Fuhrmann} L.,  et~al., 2002, \mn@doi [Publications of the Astronomical Society
  of Australia] {10.1071/AS01080}, \href
  {https://ui.adsabs.harvard.edu/abs/2002PASA...19...64F} {19, 64}

\bibitem[\protect\citeauthoryear{{Gab{\'a}nyi} et~al.,}{{Gab{\'a}nyi}
  et~al.}{2007a}]{Gabanyi-et-al-2007a}
{Gab{\'a}nyi} K.~{\'E}.,  et~al., 2007a, \mn@doi [Astronomische Nachrichten]
  {10.1002/asna.200710818}, \href
  {https://ui.adsabs.harvard.edu/abs/2007AN....328..863G} {328, 863}

\bibitem[\protect\citeauthoryear{{Gab{\'a}nyi} et~al.,}{{Gab{\'a}nyi}
  et~al.}{2007b}]{Gabanyi-et-al-2007b}
{Gab{\'a}nyi} K.~{\'E}.,  et~al., 2007b, \mn@doi [\aap]
  {10.1051/0004-6361:20067033}, \href
  {https://ui.adsabs.harvard.edu/abs/2007A&A...470...83G} {470, 83}

\bibitem[\protect\citeauthoryear{Gab{\'a}nyi, Marchili, Krichbaum, Fuhrmann,
  M{\"u}ller, Zensus, Liu  \& Song}{Gab{\'a}nyi
  et~al.}{2009}]{Gabanyi-et-al-2009}
Gab{\'a}nyi K.,  Marchili N.,  Krichbaum T.,  Fuhrmann L.,  M{\"u}ller P.,
  Zensus J.,  Liu X.,   Song H.,  2009, Astronomy \& Astrophysics, 508, 161

\bibitem[\protect\citeauthoryear{{Goldreich} \& {Sridhar}}{{Goldreich} \&
  {Sridhar}}{2006}]{Goldreich-Sridhar-2006}
{Goldreich} P.,  {Sridhar} S.,  2006, \apjl, 640, L159

\bibitem[\protect\citeauthoryear{{Heeschen}, {Krichbaum}, {Schalinski}  \&
  {Witzel}}{{Heeschen} et~al.}{1987}]{Heeschen-et-al-1987}
{Heeschen} D.~S.,  {Krichbaum} T.,  {Schalinski} C.~J.,   {Witzel} A.,  1987,
  \mn@doi [\aj] {10.1086/114583}, \href
  {http://adsabs.harvard.edu/abs/1987AJ.....94.1493H} {94, 1493}

\bibitem[\protect\citeauthoryear{{Jauncey} \& {Macquart}}{{Jauncey} \&
  {Macquart}}{2001}]{Jauncey-Macquart-2001}
{Jauncey} D.~L.,  {Macquart} J.~P.,  2001, \mn@doi [\aap]
  {10.1051/0004-6361:20010299}, \href
  {https://ui.adsabs.harvard.edu/abs/2001A&A...370L...9J} {370, L9}

\bibitem[\protect\citeauthoryear{{Jauncey}, {Kedziora-Chudczer}, {Lovell},
  {Nicolson}, {Perley}, {Reynolds}, {Tzioumis}  \& {Wieringa}}{{Jauncey}
  et~al.}{2000}]{Jauncey-et-al-2000}
{Jauncey} D.~L.,  {Kedziora-Chudczer} L.~L.,  {Lovell} J.~E.~J.,  {Nicolson}
  G.~D.,  {Perley} R.~A.,  {Reynolds} J.~E.,  {Tzioumis} A.~K.,   {Wieringa}
  M.~H.,  2000, in {Hirabayashi} H.,  {Edwards} P.~G.,   {Murphy} D.~W.,  eds,
  Astrophysical Phenomena Revealed by Space VLBI. pp 147--150

\bibitem[\protect\citeauthoryear{{Jauncey}, {Johnston}, {Bignall}, {Lovell},
  {Kedziora-Chudczer}, {Tzioumis}  \& {Macquart}}{{Jauncey}
  et~al.}{2003}]{Jauncey-et-al-2003}
{Jauncey} D.~L.,  {Johnston} H.~M.,  {Bignall} H.~E.,  {Lovell} J.~E.~J.,
  {Kedziora-Chudczer} L.,  {Tzioumis} A.~K.,   {Macquart} J.-P.,  2003, \mn@doi
  [\apss] {10.1023/B:ASTR.0000004994.54721.63}, \href
  {http://adsabs.harvard.edu/abs/2003Ap%26SS.288...63J} {288, 63}

\bibitem[\protect\citeauthoryear{{Jauncey} et~al.,}{{Jauncey}
  et~al.}{2016}]{Jauncey-et-al-2016}
{Jauncey} D.,  et~al., 2016, \mn@doi [Galaxies] {10.3390/galaxies4040062},
  \href {https://ui.adsabs.harvard.edu/abs/2016Galax...4...62J} {4, 62}

\bibitem[\protect\citeauthoryear{{Kedziora-Chudczer}}{{Kedziora-Chudczer}}{2006}]{Kedziora-Chudzer-2006}
{Kedziora-Chudczer} L.,  2006, \mn@doi [\mnras]
  {10.1111/j.1365-2966.2006.10321.x}, \href
  {http://adsabs.harvard.edu/abs/2006MNRAS.369..449K} {369, 449}

\bibitem[\protect\citeauthoryear{{Kedziora-Chudczer}, {Jauncey}, {Wieringa},
  {Walker}, {Nicolson}, {Reynolds}  \& {Tzioumis}}{{Kedziora-Chudczer}
  et~al.}{1997}]{Kedziora-Chudzer-et-al-1997}
{Kedziora-Chudczer} L.~L.,  {Jauncey} D.~L.,  {Wieringa} M.~H.,  {Walker}
  M.~A.,  {Nicolson} G.~D.,  {Reynolds} J.~E.,   {Tzioumis} A.~K.,  1997,
  \apjl, 490, L9

\bibitem[\protect\citeauthoryear{{Kedziora-Chudczer}, {Jauncey}, {Wieringa},
  {Tzioumis}  \& {Bignall}}{{Kedziora-Chudczer}
  et~al.}{2001a}]{Kedziora-Chudzer-2001a}
{Kedziora-Chudczer} L.,  {Jauncey} D.~L.,  {Wieringa} M.~A.,  {Tzioumis} A.~K.,
    {Bignall} H.~E.,  2001a, \mn@doi [\apss] {10.1023/A:1013173805039}, \href
  {http://adsabs.harvard.edu/abs/2001Ap%26SS.278..113K} {278, 113}

\bibitem[\protect\citeauthoryear{{Kedziora-Chudczer}, {Jauncey}, {Wieringa},
  {Tzioumis}  \& {Reynolds}}{{Kedziora-Chudczer}
  et~al.}{2001b}]{Kedziora-Chudzer-2001b}
{Kedziora-Chudczer} L.~L.,  {Jauncey} D.~L.,  {Wieringa} M.~H.,  {Tzioumis}
  A.~K.,   {Reynolds} J.~E.,  2001b, \mn@doi [\mnras]
  {10.1046/j.1365-8711.2001.04516.x}, \href
  {http://adsabs.harvard.edu/abs/2001MNRAS.325.1411K} {325, 1411}

\bibitem[\protect\citeauthoryear{{Kellermann} \& {Pauliny-Toth}}{{Kellermann}
  \& {Pauliny-Toth}}{1969}]{Kellermann-Pauliny-1969}
{Kellermann} K.~I.,  {Pauliny-Toth} I.~I.~K.,  1969, \mn@doi [\apjl]
  {10.1086/180305}, \href {http://adsabs.harvard.edu/abs/1969ApJ...155L..71K}
  {155, L71}

\bibitem[\protect\citeauthoryear{{Kellermann} \& {Pauliny-Toth}}{{Kellermann}
  \& {Pauliny-Toth}}{1981}]{Kellermann-Pauliny-1981}
{Kellermann} K.~I.,  {Pauliny-Toth} I. I.~K.,  1981, \mn@doi [Annual Review of
  Astronomy and Astrophysics] {10.1146/annurev.aa.19.090181.002105}, 19, 373

\bibitem[\protect\citeauthoryear{{Koay}, {Bignall}, {Macquart}, {Jauncey},
  {Rickett}  \& {Lovell}}{{Koay} et~al.}{2011}]{Koay-et-al-2011}
{Koay} J.~Y.,  {Bignall} H.~E.,  {Macquart} J.-P.,  {Jauncey} D.~L.,  {Rickett}
  B.~J.,   {Lovell} J.~E.~J.,  2011, \mn@doi [\aap]
  {10.1051/0004-6361/201117805}, \href
  {http://adsabs.harvard.edu/abs/2011A%26A...534L...1K} {534, L1}

\bibitem[\protect\citeauthoryear{{Lister} et~al.,}{{Lister}
  et~al.}{2009}]{Lister-et-al-2009}
{Lister} M.~L.,  et~al., 2009, \mn@doi [\aj] {10.1088/0004-6256/137/3/3718},
  \href {http://adsabs.harvard.edu/abs/2009AJ....137.3718L} {137, 3718}

\bibitem[\protect\citeauthoryear{{Liu} \& {Liu}}{{Liu} \&
  {Liu}}{2015}]{Liu-Liu-2015}
{Liu} J.,  {Liu} X.,  2015, \mn@doi [\apss] {10.1007/s10509-015-2393-5}, \href
  {https://ui.adsabs.harvard.edu/abs/2015Ap&SS.357..165L} {357, 165}

\bibitem[\protect\citeauthoryear{{Liu}, {Song}, {Marchili}, {Liu}, {Liu},
  {Krichbaum}, {Fuhrmann}  \& {Zensus}}{{Liu} et~al.}{2012}]{Liu-et-al-2012}
{Liu} X.,  {Song} H.-G.,  {Marchili} N.,  {Liu} B.-R.,  {Liu} J.,  {Krichbaum}
  T.~P.,  {Fuhrmann} L.,   {Zensus} J.~A.,  2012, \mn@doi [\aap]
  {10.1051/0004-6361/201219367}, \href
  {http://adsabs.harvard.edu/abs/2012A%26A...543A..78L} {543, A78}

\bibitem[\protect\citeauthoryear{{Liu}, {Liu}, {Marchili}, {Liu}, {Mi},
  {Krichbaum}, {Fuhrmann}  \& {Zensus}}{{Liu} et~al.}{2013}]{Liu-et-al-2013}
{Liu} B.-R.,  {Liu} X.,  {Marchili} N.,  {Liu} J.,  {Mi} L.-G.,  {Krichbaum}
  T.~P.,  {Fuhrmann} L.,   {Zensus} J.~A.,  2013, \mn@doi [\aap]
  {10.1051/0004-6361/201220850}, \href
  {http://adsabs.harvard.edu/abs/2013A%26A...555A.134L} {555, A134}

\bibitem[\protect\citeauthoryear{{Liu} et~al.,}{{Liu}
  et~al.}{2017}]{Liu-et-al-2017}
{Liu} X.,  et~al., 2017, \mn@doi [\mnras] {10.1093/mnras/stx1062}, \href
  {https://ui.adsabs.harvard.edu/abs/2017MNRAS.469.2457L} {469, 2457}

\bibitem[\protect\citeauthoryear{{Lovell} et~al.,}{{Lovell}
  et~al.}{2008}]{Lovell-et-al-2008}
{Lovell} J.~E.~J.,  et~al., 2008, \mn@doi [\apj] {10.1086/592485}, \href
  {http://adsabs.harvard.edu/abs/2008ApJ...689..108L} {689, 108}

\bibitem[\protect\citeauthoryear{{Macquart} \& {Jauncey}}{{Macquart} \&
  {Jauncey}}{2002}]{Macquart-Jauncey-2002}
{Macquart} J.-P.,  {Jauncey} D.~L.,  2002, \mn@doi [\apj] {10.1086/340433},
  \href {http://cdsads.u-strasbg.fr/abs/2002ApJ...572..786M} {572, 786}

\bibitem[\protect\citeauthoryear{{Macquart} \& {de Bruyn}}{{Macquart} \& {de
  Bruyn}}{2007}]{Macquart-Bruyn-2007}
{Macquart} J.-P.,  {de Bruyn} A.~G.,  2007, \mn@doi [\mnras]
  {10.1111/j.1745-3933.2007.00341.x}, \href
  {http://adsabs.harvard.edu/abs/2007MNRAS.380L..20M} {380, L20}

\bibitem[\protect\citeauthoryear{{Marchili}, {Krichbaum}, {Liu}, {Song},
  {Gab{\'a}nyi}, {Fuhrmann}, {Witzel}  \& {Zensus}}{{Marchili}
  et~al.}{2012}]{Marchili-et-al-2012}
{Marchili} N.,  {Krichbaum} T.~P.,  {Liu} X.,  {Song} H.-G.,  {Gab{\'a}nyi}
  K.~{\'E}.,  {Fuhrmann} L.,  {Witzel} A.,   {Zensus} J.~A.,  2012, \mn@doi
  [\aap] {10.1051/0004-6361/201218977}, \href
  {http://adsabs.harvard.edu/abs/2012A%26A...542A.121M} {542, A121}

\bibitem[\protect\citeauthoryear{{Marscher}}{{Marscher}}{1998}]{Marscher1998}
{Marscher} A.~P.,  1998, in {Zensus} J.~A.,  {Taylor} G.~B.,   {Wrobel} J.~M.,
  eds,  Astronomical Society of the Pacific Conference Series Vol. 144, IAU
  Colloq. 164: Radio Emission from Galactic and Extragalactic Compact Sources.
  p.~25

\bibitem[\protect\citeauthoryear{{McCallum}}{{McCallum}}{2009}]{McCallum2009}
{McCallum} J.~N.,  2009, PhD thesis, University of Tasmania

\bibitem[\protect\citeauthoryear{{McCulloch}, {Ellingsen}, {Jauncey}, {Carter},
  {Cim{\`o}}, {Lovell}  \& {Dodson}}{{McCulloch}
  et~al.}{2005}]{McCulloch-et-al-2005}
{McCulloch} P.~M.,  {Ellingsen} S.~P.,  {Jauncey} D.~L.,  {Carter} S.~J.~B.,
  {Cim{\`o}} G.,  {Lovell} J.~E.~J.,   {Dodson} R.~G.,  2005, \mn@doi [\aj]
  {10.1086/428374}, \href {http://adsabs.harvard.edu/abs/2005AJ....129.2034M}
  {129, 2034}

\bibitem[\protect\citeauthoryear{{Narayan}}{{Narayan}}{1992}]{Narayan1992}
{Narayan} R.,  1992, \mn@doi [Philosophical Transactions of the Royal Society
  of London Series A] {10.1098/rsta.1992.0090}, \href
  {http://cdsads.u-strasbg.fr/abs/1992RSPTA.341..151N} {341, 151}

\bibitem[\protect\citeauthoryear{{Nicolson}, {Glass}, {Feast}  \&
  {Andrews}}{{Nicolson} et~al.}{1979}]{Nicolson-et-al-1979}
{Nicolson} G.~D.,  {Glass} I.~S.,  {Feast} M.~W.,   {Andrews} P.~J.,  1979,
  \mn@doi [\mnras] {10.1093/mnras/189.1.29P}, \href
  {http://adsabs.harvard.edu/abs/1979MNRAS.189P..29N} {189, 29P}

\bibitem[\protect\citeauthoryear{{Ojha} et~al.,}{{Ojha}
  et~al.}{2010}]{Ojha-et-al-2010}
{Ojha} R.,  et~al., 2010, \mn@doi [\aap] {10.1051/0004-6361/200912724}, \href
  {http://adsabs.harvard.edu/abs/2010A%26A...519A..45O} {519, A45}

\bibitem[\protect\citeauthoryear{{Pen} \& {Levin}}{{Pen} \&
  {Levin}}{2014}]{Pen-Levin-2014}
{Pen} U.-L.,  {Levin} Y.,  2014, \mnras, 442, 3338

\bibitem[\protect\citeauthoryear{{Peng}, {Kraus}, {Krichbaum}  \&
  {Witzel}}{{Peng} et~al.}{2000}]{Peng-et-al-2000}
{Peng} B.,  {Kraus} A.,  {Krichbaum} T.~P.,   {Witzel} A.,  2000, \mn@doi
  [\aaps] {10.1051/aas:2000230}, \href
  {http://adsabs.harvard.edu/abs/2000A%26AS..145....1P} {145, 1}

\bibitem[\protect\citeauthoryear{{Qian}, {Quirrenbach}, {Witzel}, {Krichbaum},
  {Hummel}  \& {Zensus}}{{Qian} et~al.}{1991}]{Qian-et-al-1991}
{Qian} S.~J.,  {Quirrenbach} A.,  {Witzel} A.,  {Krichbaum} T.~P.,  {Hummel}
  C.~A.,   {Zensus} J.~A.,  1991, \aap, \href
  {https://ui.adsabs.harvard.edu/abs/1991A&A...241...15Q} {241, 15}

\bibitem[\protect\citeauthoryear{{Quirrenbach} et~al.,}{{Quirrenbach}
  et~al.}{1992}]{Quirrenbach-et-al-1992}
{Quirrenbach} A.,  et~al., 1992, \aap, \href
  {http://cdsads.u-strasbg.fr/abs/1992A%26A...258..279Q} {258, 279}

\bibitem[\protect\citeauthoryear{{Reynolds}}{{Reynolds}}{1994}]{Reynolds1994}
{Reynolds} J.~E.,  1994, ATNF Internal

\bibitem[\protect\citeauthoryear{{Richards} et~al.,}{{Richards}
  et~al.}{2011}]{Richards-et-al-2011}
{Richards} J.~L.,  et~al., 2011, \mn@doi [\apjs] {10.1088/0067-0049/194/2/29},
  \href {https://ui.adsabs.harvard.edu/abs/2011ApJS..194...29R} {194, 29}

\bibitem[\protect\citeauthoryear{{Rickett}}{{Rickett}}{1990}]{Rickett1990}
{Rickett} B.~J.,  1990, \mn@doi [\araa] {10.1146/annurev.aa.28.090190.003021},
  \href {http://cdsads.u-strasbg.fr/abs/1990ARA%26A..28..561R} {28, 561}

\bibitem[\protect\citeauthoryear{{Rickett} \& {Coles}}{{Rickett} \&
  {Coles}}{2004}]{Rickett-Coles-2004}
{Rickett} B.~J.,  {Coles} W.~A.,  2004, in American Astronomical Society
  Meeting Abstracts. p.~1539

\bibitem[\protect\citeauthoryear{{Rickett}, {Quirrenbach}, {Wegner},
  {Krichbaum}  \& {Witzel}}{{Rickett} et~al.}{1995}]{Rickett-et-al-1995}
{Rickett} B.~J.,  {Quirrenbach} A.,  {Wegner} R.,  {Krichbaum} T.~P.,
  {Witzel} A.,  1995, \aap, \href
  {http://adsabs.harvard.edu/abs/1995A%26A...293..479R} {293, 479}

\bibitem[\protect\citeauthoryear{{Rickett}, {Witzel}, {Kraus}, {Krichbaum}  \&
  {Qian}}{{Rickett} et~al.}{2001}]{Rickett-et-al-2001}
{Rickett} B.~J.,  {Witzel} A.,  {Kraus} A.,  {Krichbaum} T.~P.,   {Qian} S.~J.,
   2001, \mn@doi [\apjl] {10.1086/319493}, \href
  {http://adsabs.harvard.edu/abs/2001ApJ...550L..11R} {550, L11}

\bibitem[\protect\citeauthoryear{{Rickett}, {Kedziora-Chudczer}  \&
  {Jauncey}}{{Rickett} et~al.}{2002}]{Rickett-et-al-2002}
{Rickett} B.~J.,  {Kedziora-Chudczer} L.~L.,   {Jauncey} D.~L.,  2002, \apj,
  581, 103

\bibitem[\protect\citeauthoryear{{Rickett}, {Lazio}  \& {Ghigo}}{{Rickett}
  et~al.}{2006}]{Rickett-et-al-2006}
{Rickett} B.~J.,  {Lazio} T.~J.~W.,   {Ghigo} F.~D.,  2006, The Astrophysical
  Journal Supplement Series, 165, 439

\bibitem[\protect\citeauthoryear{{Senkbeil}, {Ellingsen}, {Lovell}, {Macquart},
  {Cim{\`o}}  \& {Jauncey}}{{Senkbeil} et~al.}{2008}]{Senkbeil-et-al-2008}
{Senkbeil} C.~E.,  {Ellingsen} S.~P.,  {Lovell} J.~E.~J.,  {Macquart} J.-P.,
  {Cim{\`o}} G.,   {Jauncey} D.~L.,  2008, \mn@doi [\apjl] {10.1086/527300},
  \href {http://adsabs.harvard.edu/abs/2008ApJ...672L..95S} {672, L95}

\bibitem[\protect\citeauthoryear{{Shabala}, {Rogers}, {McCallum}, {Titov},
  {Blanchard}, {Lovell}  \& {Watson}}{{Shabala}
  et~al.}{2014}]{Shabala-et-al-2014}
{Shabala} S.~S.,  {Rogers} J.~G.,  {McCallum} J.~N.,  {Titov} O.~A.,
  {Blanchard} J.,  {Lovell} J.~E.~J.,   {Watson} C.~S.,  2014, \mn@doi [Journal
  of Geodesy] {10.1007/s00190-014-0706-z}, \href
  {http://adsabs.harvard.edu/abs/2014JGeod..88..575S} {88, 575}

\bibitem[\protect\citeauthoryear{{Shepherd}}{{Shepherd}}{1997}]{Shepherd1997}
{Shepherd} M.~C.,  1997, in {Hunt} G.,  {Payne} H.,  eds,  Astronomical Society
  of the Pacific Conference Series Vol. 125, Astronomical Data Analysis
  Software and Systems VI. p.~77

\bibitem[\protect\citeauthoryear{{Sijbers}, {Scheunders}, {Bonnet}, {Van Dyck}
  \& {Raman}}{{Sijbers} et~al.}{1996}]{Sijbers-et-al-1996}
{Sijbers} J.,  {Scheunders} P.,  {Bonnet} N.,  {Van Dyck} D.,   {Raman} E.,
  1996, \mn@doi [Magnetic Resonance Imaging]
  {https://doi.org/10.1016/S0730-725X(96)00219-6}, 14, 1157

\bibitem[\protect\citeauthoryear{{Simonetti}, {Cordes}  \&
  {Heeschen}}{{Simonetti} et~al.}{1985}]{Simonetti-et-al-1985}
{Simonetti} J.~H.,  {Cordes} J.~M.,   {Heeschen} D.~S.,  1985, \mn@doi [\apj]
  {10.1086/163418}, \href {http://cdsads.u-strasbg.fr/abs/1985ApJ...296...46S}
  {296, 46}

\bibitem[\protect\citeauthoryear{{Spangler}, {Fanti}, {Gregorini}  \&
  {Padrielli}}{{Spangler} et~al.}{1989}]{Sprangler-et-al-1989}
{Spangler} S.,  {Fanti} R.,  {Gregorini} L.,   {Padrielli} L.,  1989, \aap,
  \href {http://cdsads.u-strasbg.fr/abs/1989A%26A...209..315S} {209, 315}

\bibitem[\protect\citeauthoryear{{Stickel}, {Fried}  \& {Kuehr}}{{Stickel}
  et~al.}{1989}]{Stickel-et-al-1989}
{Stickel} M.,  {Fried} J.~W.,   {Kuehr} H.,  1989, \aaps, \href
  {http://adsabs.harvard.edu/abs/1989A%26AS...80..103S} {80, 103}

\bibitem[\protect\citeauthoryear{{Turner}, {Ellingsen}, {Shabala}, {Blanchard},
  {Lovell}, {McCallum}  \& {Cimò}}{{Turner} et~al.}{2012}]{Turner-et-al-2012}
{Turner} R.~J.,  {Ellingsen} S.~P.,  {Shabala} S.~S.,  {Blanchard} J.,
  {Lovell} J.~E.~J.,  {McCallum} J.~N.,   {Cimò} G.,  2012, \apjl, 754, L19

\bibitem[\protect\citeauthoryear{{V{\'e}ron-Cetty} \&
  {V{\'e}ron}}{{V{\'e}ron-Cetty} \& {V{\'e}ron}}{2006}]{Veron-Cetty-2006}
{V{\'e}ron-Cetty} M.-P.,  {V{\'e}ron} P.,  2006, \mn@doi [\aap]
  {10.1051/0004-6361:20065177}, \href
  {http://adsabs.harvard.edu/abs/2006A%26A...455..773V} {455, 773}

\bibitem[\protect\citeauthoryear{{Wagner} \& {Witzel}}{{Wagner} \&
  {Witzel}}{1995}]{Wagner-Witzel-1995}
{Wagner} S.~J.,  {Witzel} A.,  1995, \mn@doi [\araa]
  {10.1146/annurev.aa.33.090195.001115}, \href
  {http://adsabs.harvard.edu/abs/1995ARA%26A..33..163W} {33, 163}

\bibitem[\protect\citeauthoryear{{Walker}}{{Walker}}{1998}]{Walker1998}
{Walker} M.~A.,  1998, \mn@doi [\mnras] {10.1046/j.1365-8711.1998.01238.x},
  \href {http://adsabs.harvard.edu/abs/1998MNRAS.294..307W} {294, 307}

\bibitem[\protect\citeauthoryear{{Walker}}{{Walker}}{2001}]{Walker2001}
{Walker} M.~A.,  2001, \mn@doi [\mnras] {10.1046/j.1365-8711.2001.04104.x},
  \href {http://adsabs.harvard.edu/abs/2001MNRAS.321..176W} {321, 176}

\bibitem[\protect\citeauthoryear{{Walker}, {De Bruyn}  \& {Bignall}}{{Walker}
  et~al.}{2009}]{Walker-et-al-2009}
{Walker} M.~A.,  {De Bruyn} A.~G.,   {Bignall} H.~E.,  2009, \mn@doi [Monthly
  Notices of the Royal Astronomical Society]
  {10.1111/j.1365-2966.2009.14942.x}, 397, 447

\bibitem[\protect\citeauthoryear{{Walker}, {Tuntsov}, {Bignall}, {Reynolds},
  {Bannister}, {Johnston}, {Stevens}  \& {Ravi}}{{Walker}
  et~al.}{2017}]{Walker-et-al-2017}
{Walker} M.~A.,  {Tuntsov} A.~V.,  {Bignall} H.~E.,  {Reynolds} C.,
  {Bannister} K.~W.,  {Johnston} S.,  {Stevens} J.,   {Ravi} V.,  2017, \apj,
  843, 15

\makeatother
\end{thebibliography}









\end{document}